\documentclass{aa}
\usepackage{graphicx}
\usepackage{txfonts}
\usepackage{longtable}
\usepackage{natbib}

\begin{document} 


\title{Limits on the Macho Content of the Galactic Halo from the  EROS-2
Survey of the Magellanic Clouds\thanks{
Based on observations made with the Marly
  telescope at the European Southern Observatory, La Silla, Chile.}}
\author{
P.~Tisserand\inst{1}\thanks{Now at
Research School of Astronomy and Astrophysics,
Australian National University,
Mount Stromlo Obs.,
Cotter Rd., Weston, ACT 2611, Australia},
L. Le Guillou\inst{1}\thanks{now at LPNHE, CNRS-IN2P3 and Universit\'es 
Paris 6 et Paris 7, 4 Place Jussieu,
F-75252 Paris Cedex 05 France},
C.~Afonso\inst{1}\thanks{Now at Max-Planck-Institut f\"ur Astronomie,
Koenigstuhl 17, D-69117 Heidelberg, Germany
},
J.N.~Albert\inst{2},
J.~Andersen\inst{5},
R.~Ansari\inst{2},
\'E.~Aubourg\inst{1}\thanks{Also at APC, 10, 
rue Alice Domon et Léonie Duquet, F-75205 Paris Cedex 13, France},
P.~Bareyre\inst{1},
J.P.~Beaulieu\inst{3},
X.~Charlot\inst{1},
C.~Coutures\inst{1,3},
R.~Ferlet\inst{3},
P. Fouqu\'e\inst{7,8},
J.F.~Glicenstein\inst{1},
B.~Goldman\inst{1}\thanks{Now at  Max-Planck-Institut f\"ur Astronomie,
Koenigstuhl 17, D-69117 Heidelberg, Germany},
A.~Gould\inst{6},
D.~Graff\inst{6}\thanks{Now at Division of Medical Imaging Physics,
Johns Hopkins University
Baltimore, MD 21287-0859, 
USA},
M.~Gros\inst{1},
J.~Haissinski\inst{2},
C. Hamadache\inst{1},
J.~de Kat\inst{1},
T.~Lasserre\inst{1},
\'E.~Lesquoy\inst{1,3},
C.~Loup\inst{3},
C.~Magneville\inst{1},
J.B.~Marquette\inst{3},
\'E.~Maurice\inst{4},
A.~Maury\inst{8}\thanks{Now at San Pedro de Atacama Celestial Exploration,
Casilla 21, San Pedro de Atacama, Chile},
A.~Milsztajn \inst{1},
M.~Moniez\inst{2},
N.~Palanque-Delabrouille\inst{1},
O.~Perdereau\inst{2},
Y.R. Rahal\inst{2},
J.~Rich\inst{1},
M.~Spiro\inst{1},
A.~Vidal-Madjar\inst{3},
L.~Vigroux\inst{1,3},
S.~Zylberajch\inst{1}
\\   \indent   \indent
The EROS-2 collaboration
}    
\institute{
CEA, DSM, DAPNIA,
Centre d'\'Etudes de Saclay, F-91191 Gif-sur-Yvette Cedex, France
\and
Laboratoire de l'Acc\'{e}l\'{e}rateur Lin\'{e}aire,
IN2P3 CNRS, Universit\'e de Paris-Sud, F-91405 Orsay Cedex, France
\and
Institut d'Astrophysique de Paris,
UMR 7095 CNRS, Universit\'e Pierre \& Marie Curie,
98~bis Boulevard Arago, F-75014 Paris, France
\and
Observatoire de Marseille,
2 place Le Verrier, F-13248 Marseille Cedex 04, France
\and
The Niels Bohr Institute, Copenhagen University, Juliane Maries Vej 30,
DK-2100 Copenhagen, Denmark
\and
Department of Astronomy, Ohio State University, Columbus,
OH 43210, U.S.A.
\and
Observatoire Midi-Pyr\'en\'ees, Laboratoire d'Astrophysique (UMR 5572), 
14 av. E. Belin, F-31400 Toulouse, France
\and
European Southern Observatory (ESO), Casilla 19001, Santiago 19, Chile
}

\date{Received;accepted}

\authorrunning{P. Tisserand et al.}
\titlerunning{Limits on the Macho content of the Galactic Halo from EROS-2}

\def\lsim{{\lesssim}}
\def\au{{\rm AU}}
\def\tempest%
{\begin{array}{ccc}
1 & 1 & 1 \\
1 & 1 & 1 \\
4 & 3 & 8
\end{array}}
\def\gsim{{{}_>\atop{}^{{}^\sim}}}
\def\lsim{{{}_<\atop{}^{{}^\sim}}}
\def\kms{{\rm km}\,{\rm s}^{-1}}
\def\kpc{{\rm kpc}}
\def\sec{{\rm s}}
\def\e{{\rm E}}
\def\rel{{\rm rel}}
\def\btheta{{\vec\theta}}
\def\bmu{{\vec\mu}}
\def\bpi{{\vec\pi}}
\def\Teff{{T_{\rm eff}}} 
\def\msun{\rm M_\odot} 
\def\deg{\rm deg}
\def\day{\rm d}
\def\te{t_{\rm E}}
\def\re{r_{\rm E}}
\def\reros{R_{\rm eros}}
\def\beros{B_{\rm eros}}
\def\Vj{V_{\rm J}}
\def\Rc{R_{\rm C}}
\def\Ic{I_{\rm C}}

\abstract
{} 
{
The EROS-2 project was designed to test the hypothesis that massive
compact halo objects (the so-called ``machos'') could be a major component
of the dark matter halo of the Milky Way galaxy. To this end, EROS-2
monitored over 6.7 years
$33\times10^6$ stars in the Magellanic clouds for microlensing
events caused by such objects.
}
{
In this work, we use only a subsample of $7\times10^6$  bright stars
spread over $84\,\deg^2$ of the LMC and $9\,\deg^2$ of the SMC.
The strategy of using only bright stars
helps to discriminate against background events due to variable stars and
allows a simple determination of the effects of
source confusion (blending).
The use of a large solid angle makes the survey relatively
insensitive to effects that could make the optical depth 
strongly direction dependent.
}
{
Using this sample of bright stars, only one candidate event was
found, whereas $\sim39$ events would have been expected if
the Halo were entirely populated by objects of
mass $M\sim0.4M_{\odot}$.
Combined with the results of EROS-1, 
this  implies that the optical depth toward the
Large Magellanic Cloud (\object{LMC}) 
due to such lenses is $\tau<0.36\times10^{-7}$ (95\%CL), corresponding
to a fraction of the halo mass of less than 8\%. 
This optical depth is considerably less than that
measured by the MACHO collaboration in the central region
of the LMC.
More generally, machos in the
mass range $0.6\times10^{-7}M_\odot<M<15M_{\odot}$ are ruled out
as the primary occupants of the Milky Way Halo.
}  
{}


\maketitle           

\section{Introduction} 
\label{section:introduction}

Since  the proposal \citep{Pac1986, Pet81} that dark matter in 
the form of faint compact objects 
(machos\footnote{for
``Massive Astrophysical Compact Halo Objects''
\citep{GRI91}, not to be confused with the ``MACHO collaboration.''}) 
could be found through
gravitational microlensing, 
the EROS, MACHO, OGLE,  MOA and SuperMACHO collaborations  have
monitored millions of stars in the
Magellanic Clouds  to search for microlensing events.
Such events  would be due to  
a lensing object passing near the line of sight
toward a background Magellanic star, causing a 
transient magnification of the star's primary image
as well as creating a secondary image.
Neither the image separation nor the
image size are normally resolvable, so the only easily observable
effect during an event 
is an apparent transient amplification of the star's flux.
The amplification is greater than a factor 1.34 if the
line of sight to the star passes within the lens's Einstein ring of
squared radius 
$\re^2=4GMD_{\rm s}x(1-x)/c^2$ where  $D_{\rm s}$ is the distance
to the star and  $xD_{\rm s}$ is the distance to the lens of mass $M$.
The optical depth for microlensing, 
i.e. the probability that 
at a given time a given star  
is amplified by more than a factor
1.34, is
\begin{equation}
\tau \;=\; \frac{4\pi G D_{\rm s}^2}{c^2}\int_0^1 dx \rho(x)x(1-x) 
\;,
\label{optdepthint}
\end{equation}
where $\rho$ is the mass density of lenses.
For source stars in the Magellanic Clouds, the order of magnitude of $\tau$
is $ fv_{\rm rot}^2/c^2 \sim f\times10^{-6}$ where
$v_{\rm rot}\sim 220\,\kms$ is the rotation velocity
of the Milky Way and $f$ is the fraction of the halo mass
that is comprised of lensing objects.
The factor of proportionality between $\tau$ and $fv_{\rm rot}^2/c^2$  
depends on the structure of the Halo. 
The benchmark value is often taken to be that
for a spherical isothermal halo of core radius $5\,\kpc$,
the so-called ``S model'' used by the MACHO
collaboration \citep{macho57,GRI91}.
For the Large Magellanic
Cloud (LMC) this gives 
\begin{equation}
\tau_{\rm lmc}=4.7f\times10^{-7} \;.
\end{equation}
For the Small Magellanic Cloud  (\object{SMC}), the S model gives
$\tau_{\rm smc}\sim1.4\tau_{\rm lmc}$.
For a flattened halo, one finds a smaller value, typically
$\tau_{\rm smc}\sim\tau_{\rm lmc}$ 
\citep{sakgould}.

Magellanic stars can also be lensed by non-halo
stars, either in the Magellanic Clouds or in the Milky Way
disk. Lensing by disk stars is expected to have
an optical depth of order $10^{-8}$ \citep{macho57}.  
The optical depth for lensing by ``self-lensing'', i.e. lensing
by stars in the Clouds, 
is expected to range from $\sim5\times10^{-8}$ in the
center of the LMC bar to $\sim0.5\times10^{-8}$ at $3\,\deg$
from the bar \citep{jetzerself}.
For the SMC, the self-lensing optical depth
is expected to be somewhat larger,   $\sim4\times10^{-8}$ 
averaged over the central $10\,\deg^2$ \citep{graffsmc}.

Microlensing events are characterized by a timescale $\te$
giving the time for the lens to travel a distance corresponding
to  its Einstein radius, $\te=\re/v_t$ where $v_t$ is the 
lens's transverse  velocity relative to the line of sight.  
For high amplification events, $2\te$
is the time over which the amplification is $A>1.34$.
Since $\re$ is proportional to the square root of the
lens mass $M$, the mean $\te$ will scale like $M^{1/2}$.
The  S model has a 3-dimensional macho velocity dispersion
of $270\,\kms$ and gives
\begin{equation}
\langle \te\rangle
\;\sim\; 70\left(\frac{M}{M_{\odot}}\right)^{1/2}\,{\rm days} \;.
\label{smodelteeq}
\end{equation}

Much excitement was generated by  the MACHO collaboration's
measurement of the LMC microlensing rate which
suggested that a significant amount of the Milky Way's Halo 
is comprised of machos.
Their latest analysis  \citep{macho57} used 13/17 observed 
events\footnote{13 of the 17 events satisfied their so-called A criteria
intended  to identify high signal-to-noise events.  The other
4 events, so-called B events,  are selected by looser cuts.}
to measure an optical depth of 
$\tau_{\rm lmc}/10^{-7}=1.2^{+0.4}_{-0.3}\;(stat)$ with an additional
20\% to 30\% systematic error.  This would correspond to
a Halo fraction $0.08<f<0.50$ (95\% CL).
The mean $\te$ of their events was $40\,\day$ corresponding
to machos  in the mass range $0.15M_\odot<M<0.9M_\odot$.
On the other hand
the EROS collaboration \citep{las00,EROSMC} has placed only an upper
limit on the halo fraction, $f<0.2$ (95\% CL) for objects in this mass range,
ruling out a large part of the  range of $f$ favored by the
MACHO collaboration.

\citet{bennetttau} argued that the MACHO optical depth should be reduced
to $\tau_{\rm lmc}/10^{-7}=1.0\pm0.3$ in order to take into account
contamination by variable stars.  This paper made use of
the observation by the EROS collaboration \citep{tis04} of further
variability of one of the MACHO A candidates, indicating intrinsic stellar
variability.
The paper also noted that the spectrum  of the MACHO B
candidate MACHO-LMC-22 indicated that the source is
an active background galaxy, 
as reported in \citet{machobigmass} where the event was eliminated
from the sample for studying high-mass lenses.
Using four MACHO A candidates whose microlensing nature was
confirmed by precision photometry and the one A candidate rejected
as a variable star,
\citet{bennetttau} performed a likelihood analysis to argue that
$11\pm1$ of the 13 A candidates are likely to be microlensing events, yielding
the revised optical depth.

Machos can also be searched for by monitoring M31 and looking for
temporal variations of surface brightness consistent with a star
in M31 being microlensed.
Candidate events have been reported by the
VATT \citep{Ugle04},
WeCAPP \citep{Riff03},
POINT-AGAPE  \citep{PH2005}, 
MEGA \citep{mega2005} and
Nainital \citep{nainital} collaborations.
The POINT-AGAPE and MEGA collaborations presented efficiency calculations
allowing them to constrain the content of the  M31 and Milky Way halos.
The disagreement between these two collaborations parallels that between
the MACHO and EROS collaborations with the AGAPE collaboration
finding a halo fraction in the range $0.2<f<0.9$, 
while the MEGA collaboration finds a halo fraction $f<0.3$.

In this paper, we extend our previous analysis to find
$\tau_{\rm lmc}<0.36\times 10^{-7}$ (95\% CL) for $M\sim 0.4M_\odot$,
corresponding to $f<0.08$. 
Unlike the previous EROS limit, this is significantly
lower than the optical depth measured by the MACHO collaboration.  
Unlike all previous analyses,
we use only
a bright, well-measured subsample of the Magellanic stars, 
about 20\% of the total.  We believe that the use of this
bright subsample gives more reliable limits on the
optical depth than measurements using faint  stars.  
There are two reasons for this.  
First, bright stars have well reconstructed
light curves that permit discrimination of intrinsically variable
stars.  
Second, the use of bright stars makes it relatively simple to  estimate
so-called blending effects 
where reconstructed fluxes can receive contributions from more
than one star, complicating the interpretation of events.

EROS-2 is a second generation microlensing experiment.  The
first generation, EROS-1, consisted of two programs, both
at the European Southern Observatory (ESO) at La Silla, Chile.
The first program \citep{erosplates} used Schmidt photographic plates 
to monitor a $27\,\deg^2$ region containing the  LMC bar 
during the southern summer from October, 1990 through April, 1993.
With a sampling frequency of up to one image per night, it was sensitive
mostly to machos in the range $10^{-4}M_\odot<M<1M_\odot$.
The second program \citep{renault98} used
a $0.4\,\deg^2$ CCD mosaic from December 1991 through March, 1995
to monitor one field in the LMC bar
and another in the SMC.  With up to 40 images taken per night,
this program was sensitive 
mostly to machos in the range $10^{-7}M_\odot<M<10^{-3}M_\odot$.
The results of these two EROS-1 programs are summarized in \citet{renault97}.

The second generation program described here, EROS-2,
used the Marly 1 meter telescope
at ESO, La Silla.                                
The telescope was equipped with two
$0.95\,\deg^2$ CCD mosaics to monitor $93\,\deg^2$ in the Magellanic Clouds,
$63\,\deg^2$ in the Galactic Bulge, and $28\,\deg^2$ in the spiral arms
of the Milky Way.  
The observations were performed between 
July 1996 and February 2003 (JD between 2,450,300 and 2,452,700). 

Besides the Magellanic results presented here,
EROS-2 has also published measurements of the optical depth
toward the Galactic Bulge \citep{Afonso,hamaa}.
The measured optical depth is in agreement with
that measured by MACHO \citep{machocgpop}, and OGLE-2  \citep{sumiogle}
and with the predictions of
Galactic models \citep{EVA02,Bissantz,HanGould03,woodmao}.

EROS-1 and EROS-2 overlapped with the MACHO program that
monitored $\sim13.4\,\deg^2$ of the LMC from July, 1992
through January, 2000.  Three other wide-angle microlensing 
searches are now in operation:  
MOA\footnote{http://www.phys.canterbury.ac.nz/moa/}  (since August, 1998),
OGLE-3\footnote{http://bulge.astro.princeton.edu/$\sim$ogle/} (since June, 2001)
and SuperMACHO\footnote{http://www.ctio.noao.edu/$\sim$supermacho/} 
(since October, 2001).

In this article, we report on the analysis of the full EROS-2
data set (July 1996 till February 2003) toward the Magellanic Clouds. 
Our previous analyses reported on 5 years of data and 5 million stars 
toward the SMC \citep{EROSMC}, and 2 years of data and 17 million
stars toward the LMC \citep{las00}. An update of the LMC 
analysis was reported in \citet{mil00} and \citet{lasthesis}, 
which dealt with 25 million
stars and 3 years.
The limits determined in the previous analyses are refined in the analysis
presented here. 
More details about the present
analysis can be found in \citet{tis04}.

The plan of the article is as follows.
In Section \ref{setupsec}, we recall the basics of the EROS-2 setup, give the
general characteristics of the data sample and describe the data
reduction steps used to produce the light curves. 
Section \ref{brightsec} presents the Bright-Star Sample of stars to be used
in the measurement of the optical depth.
In Section \ref{analsec}, we describe the 
selection criteria used to choose the microlensing candidates
in the Bright-Star Sample.
(Selection of events in the full sample is described in 
Appendix \ref{fullselectionsec}.) 
Section \ref{candsec} presents the final sample of selected events
from the Bright-Star Sample
as well as  events found by relaxing the selection criteria.
In Section \ref{formersec} we
discuss the status of former EROS-1 and EROS-2 microlensing
candidates as well as those of the MACHO collaboration.
In Section \ref{effsec}, 
we describe the computation of the EROS-2 detection efficiency.
Section \ref{limitssec} 
presents the limit on the 
optical depth and on the
abundance of machos in the Galactic halo
by combining all EROS-1 and EROS-2 data. 
We conclude in Section \ref{discussionsec} with a  
discussion of  the significance of the
limit, and a comparison with the results of \citet{macho57}.

\section{Experimental setup, observations and data reduction}
\label{setupsec}

The EROS-2 Marly telescope, camera, 
telescope operations and data reduction are 
described in \citet{bau97}, \citet{smc1}, and references therein. 
Here we give only general information, and details or 
modifications that are specific to the present analysis.

\subsection{The setup and the data}

The Marly telescope is a one meter diameter Ritchey-Chr\'etien 
(f~=~5.14m), equipped with two wide angle CCD cameras.
Each camera is a mosaic of 8 CCDs, 2 along right ascension and 4
along declination. Each CCD has 2048~x 2048 pixels of 
$15 \times 15\, \mu$m$^2$ size, corresponding to 
$0.602 \times 0.602$ arcsec$^2$.
Images were taken simultaneously in two wide passbands, 
so-called $\reros$ centered 
close to the $I_C$ standard band, and $\beros$ 
intermediate between the standard V and R bands. 
While no  results presented here depend on the photometric
calibration, almost all of our fields could be calibrated using
stars from the  catalogs of the 
Magellanic Clouds Photometric Survey  \citep{zaritsky}.  For $4.5\,\deg^2$,
the calibration was checked with the  
OGLE-II  catalog \citep{ogleIIphotometry}.  
To a precision of $\sim0.1\,{\rm mag}$,
the EROS magnitudes satisfy
\begin{equation}
\reros=\Ic \hspace*{5mm} \beros\;=\;\Vj - 0.4(\Vj-\Ic) \;.
\end{equation}

\begin{figure}
 \centering
\includegraphics[width=7.8cm]{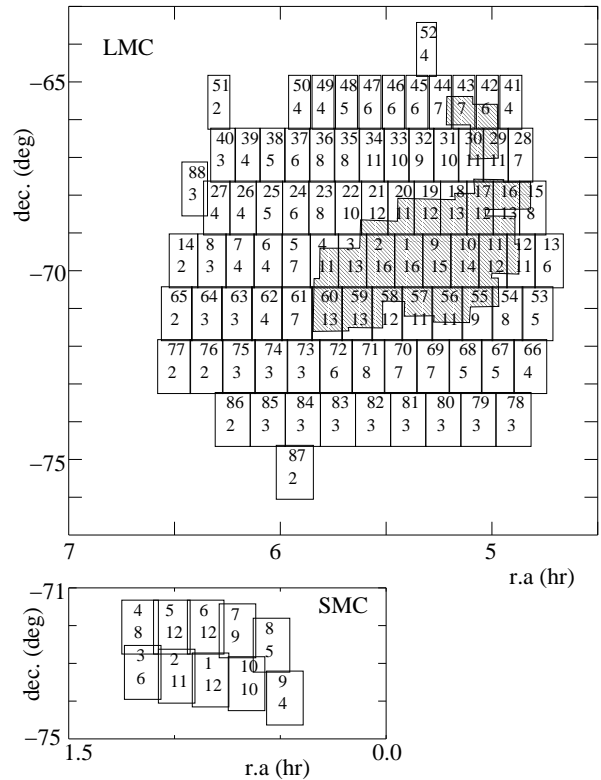}
 \caption{
Map of the EROS-2 LMC  and SMC fields in equatorial coordinates.
A total of 88 LMC and 10 SMC 
fields were monitored.  The first number in each field is the field
number and the second is the number of bright stars 
(as defined in Section \ref{brightsec})
in the field 
in units of  $10^4$.
The two shaded regions (the larger one centered on the LMC bar)
are the $13.4\,\deg^2$ 
used by the MACHO collaboration to measure the optical
depth \citep{macho57}.
}
\label{fieldsfig}
\end{figure}

The observed fields correspond to 0.95 deg$^2$ each and they are 
twice as large in declination than in right ascension
($1.38 \times 0.69$ deg$^2$). A total of 88 fields have been monitored 
toward the LMC and 10 toward the SMC. 
The positions of the fields are shown in Figure \ref{fieldsfig}.
The exposure times ranged from $180\,{\rm s}$ 
to $900\,{\rm s}$.  The fields have been 
sampled differently, according to their stellar density and distance 
from the optical centers of the LMC and the SMC.
Broadly, there are three families of LMC fields~: in the first 3 
years of operation, 22 outer fields were seldom imaged, 22
fields were imaged about 100 times and the remaining 44 inner fields
were imaged over 200 times.  
Later, from July 1999 on, all 88 LMC fields were imaged with a similar 
sampling. The number of photometric measurements, per star and per band,
ranges from 300 to 600 in the LMC fields. 

The ten SMC fields were imaged with a similar sampling, except in
2001 and 2002 when the inner six fields were imaged twice 
as often as the four outer ones. The number of photometric measurements, 
per star and per band, ranges from 600 to 900 in these SMC fields.

After rejection of bad images (11\%) 
due to bad seeing, high sky 
background or instrumental problems, 
the average numbers of measurements used in the analysis, 
per star and per band, are about 430 and 780 
in the LMC and SMC fields respectively.

\subsection{Template Images}

The production of light curves proceeded in three steps~: template 
image construction, star catalog production from the templates, 
and photometry of individual images to obtain the light curves.

The template images were obtained by co-adding 15 images of each field;
they were resampled so that the templates have
twice as many pixels as the original images. 
The 15 images were chosen among the best ones available, i.e. with low sky 
background, good seeing, and with a large number of stars
(as estimated from a quick first look algorithm). All images susceptible to 
enter the template construction were checked for the absence of long tracks, 
caused by satellites, planes or meteors. For technical reasons
linked to computing and to PSF variation within one CCD, 
the CCD images were divided in four quadrants, such that
there were in total 6272 template images (98 fields, 8 CCDs, 4
quadrants, 2 passbands). In order to ensure relatively uniform
zeropoints of the 6272 templates, we required that
the first image used in template construction
(to which the other 14 images were photometrically aligned)
be registered within a short 
time interval with good and uniform sky conditions (23rd to 26th, 
November 2000). For each field,  the same epochs were used in the
construction of the templates in the two passbands. 

The stars were identified on the template images using a pseudo-image that
we call a correlation image. Each pixel of this image contains the 
correlation coefficient 
of neighboring pixels 
of the template itself with a two-dimensional
Gaussian PSF. Each group of pixels satisfying some threshold value 
on the pseudo-image was retained as a star in the catalog. 
In previous EROS-2 analyses, 
the thresholds were identical for all templates; this had led to over 20\%
failures in this cataloging step. The present analysis has chosen to
progressively relax the thresholds when such failures occur; in this way,
the cataloging step failures were drastically reduced. (The number of 
identified stars on fields using relaxed thresholds was lower on average.)
The star catalogs 
obtained from the template images in the two passbands were then merged.
A star was retained only if it was detected in both. The overall efficiency
of the template plus star catalog construction was excellent; only 24 CCD
quadrants could not be processed out of a total of 3136.

Examples of color-magnitude diagrams can be found in 
Figures \ref{smc1fig} to \ref{macho23fig} and \ref{lm055fig} to \ref{lm085fig}.
They are all characterized by a prominent group of clump giants
and a main sequence whose relative strength varies from field to field.
There are also stars that are redder
than Magellanic red giants.  Most   of them are  likely to be foreground
stars in the disk of the Milky Way and  their number
is consistent with the predictions of Galactic models \citep{besancon}.

\subsection{Light Curves}

\begin{figure}
 \centering
\includegraphics[width=7.8cm]{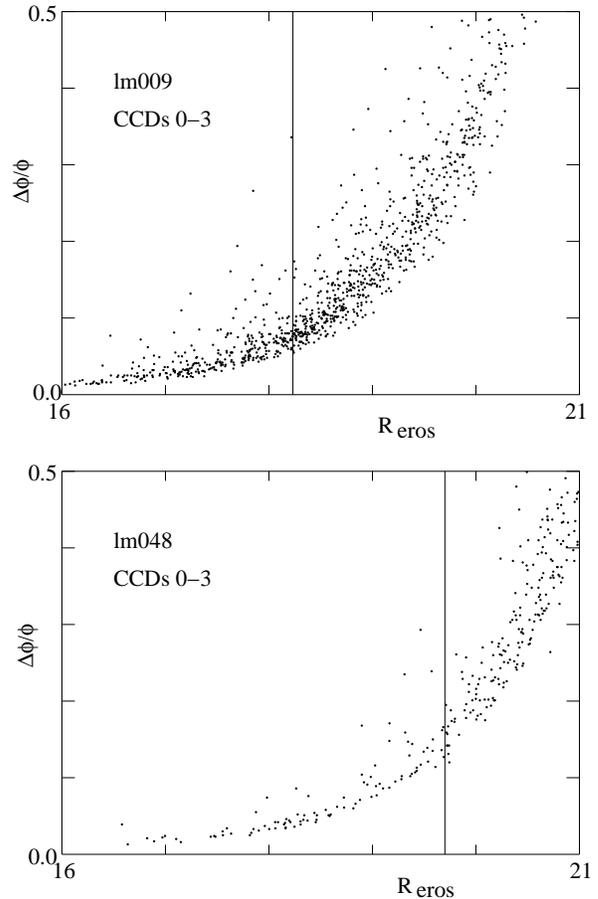}
 \caption{
The photometric precision as a function of $\reros$ for the
dense field lm009 (top) and the sparse field lm048 (bottom).
Each point represents  the r.m.s. dispersion
 of the measurements for a single star
after elimination of outliers.
The vertical line shows the position of the bright-star magnitude
cut (\ref{brightdefeq}) for the field, 
$\reros=18.23$ for lm009 CCDs 0-3
and $\reros=19.7$ for lm048 CCDs 0-3.
}
\label{lm009048resfig}
\end{figure}

Photometry was then performed on each image of a given quadrant in turn
with soft\-ware spe\-cific\-ally
designed for
crowded fields, PEIDA (Photo\-m\'etrie et \'Etude d'Images Destin\'ees
\`a l'Astrophysique) \citep{ANS96}.
First, the image was geometrically aligned with the template. Then,
imposing the star position determined from the template image, 
PSF-fitting
photometry was performed for all stars by a linear least-squares method
involving the star and all neighbors closer than 11 pixels, plus a sky
background. An estimate of the error on this flux measurement was computed 
that depends on the photon statistics and on the overall quality of the 
photometered image. Typical photometric precisions are shown in Figure
\ref{lm009048resfig}.

Before the analysis was started, we removed from the light curves measurements
taken under far from normal conditions. This happened not infrequently, as the
data taking policy was to work whenever possible, leaving to the analysis
the task of rejecting these abnormal measurements. These were identified by 
extreme values of the sky background, seeing or absorption. In addition,
images where the photometry failed for over 40\% of cataloged stars were 
eliminated, as well as images for which more than 12\% of the stars showed an
excursion from their average flux larger than three standard deviations.
Depending on the CCD, the number of rejected images varied between 
7 and 18\%, with an average of 11\%.\footnote{
The largest single cause for rejection 
was the malfunction that affected 5 CCDs of the $\reros$ passband camera 
starting in January 2002. We have thus chosen to reject all measurements 
taken in 2002 and 2003 for these 5 CCDs.}

To reduce systematic errors in the photometry,
each light curve was searched for significant
linear correlations between the measured flux and three observational 
variables, the seeing, the hour angle and the airmass. This was done 
independently in the two passbands.  
The measured fluxes were corrected linearly by requiring a vanishing
correlation coefficient between the corrected fluxes and the given 
variable. 
The largest correlation was found with the seeing, in both passbands. 
For the bright stars considered in this work,
the correction has only a small influence
on the point-to-point dispersion of the lightcurves,
reducing it on average by 6\%.

A potential problem with this correction is that an artificial correlation
can be induced if a real flux variation happens to occur during
a period of poor seeing (for example).  In this case, application
of the correction would reduce the amplitude of the real flux variation
but increase the point-to-point dispersion of the curve by making an
incorrect flux correction.
To guard against this possibility, we did not apply the correction if
it increases the
point-to-point dispersion of the light curves.

\subsection{The Full Sample of Stars}

A total of  58.4 million objects 
were found on the template images of both passbands
--~51.8 million in the LMC and 6.6 million in
the SMC. 
We chose to ignore 
those stars for which the association between the objects 
detected separately in the two passbands was doubtful or ambiguous 
(8.4 million) and objects that are dimmer than about twice the typical 
size of sky background fluctuations (14.9 million). We rejected light curves
for which more than half of the photometry points are absent (1.0 million).
Finally, we did not consider stars whose photometry is unstable due to
its environment. This includes stars close to a very bright field star 
($\Vj < 10.1$, probably in the Galactic disk) and stars close to a visible 
diffraction feature in the PSF of bright stars; these two categories 
contain respectively 1.1\% and 0.8\% of the remaining stars.

We removed from consideration stars in field lm003, CCD0 which
has numerous events caused by light echos from SN1987a. The
echos generate arc-like images
that appear to move a few arcsec per year\footnote{Using 
1000 pictures of this region co-added by
groups of 10, we have produced a film that shows the motion of the light
echos; it can be found at the URL
{\tt http://eros.in2p3.fr/EchoesSN1987a/}},
causing false variations of a star's flux as the arc passes 
through the star's position.

After these cuts, 33.4 million objects remained, 29.2 million in the LMC 
and 4.2 million in the SMC.  This constitutes the Full Sample of EROS-2 stars.
In the next section we describe the selection of the 
Bright-Star Sample used 
for the measurement of optical depths.

\section{The Bright-Star Sample}
\label{brightsec}

We have chosen to restrict our analysis to the  Bright Sample of
stars defined below.  Besides the obvious advantage of ensuring
a good photometric resolution, we do this to simplify the evaluation
of the number of expected events predicted by a model.
In crowded fields, this evaluation is complicated by ``blending'', i.e. 
the fact that photometry of a given \emph{object} 
can be influenced by more than one
\emph{star}.  We shall see that these effects are rather small and simple
to evaluate for the Bright-Star Sample  (which should, strictly speaking,
be called the Bright-Object Sample).

Limiting the number of studied objects limits the sensitivity
of the experiment so the magnitude cut must be a compromise
between quantity and quality of objects.
The efficiency calculations of Section \ref{effsec} for unblended sources
indicated  that keeping only the $\sim20\%$ brightest
stars reduces by only $\sim50\%$ the number of simulated microlensing events
that pass our selection criteria.
We therefore initially 
decided to accept a nominal $\sim50\%$ loss of events
by requiring $\reros<R_{\rm median}$, where $R_{\rm median}$ is
the median $\reros$ for simulated unblended microlensing events passing
our event selection criteria.  $R_{\rm median}$ ranges from
$\sim18.2$ in the densest fields to $\sim 20.5$ in the sparsest.
However, 
for sparse fields far from the LMC bar, 
in order to have a reasonable object reconstruction efficiency,
a stricter cut was found to be  necessary,
$\reros<19.7$.

The final sample of bright stars is therefore defined by
\begin{equation}
16.0\;<\; R_{\rm eros}\;<\; R_{\rm max} \hspace*{5mm} 
R_{\rm max}=min(R_{\rm median},19.7) \;.
\label{brightdefeq}
\end{equation}
The minimum magnitude $\reros=16$ was chosen to avoid the
large number of variable stars brighter than this.

The position of the magnitudes cut are shown in the color-magnitude
diagrams of 
Figures \ref{smc1fig} to \ref{macho23fig} and \ref{lm055fig} to \ref{lm085fig}.
Generally speaking, the cut includes clump giants but not
the numerous main sequence stars seen far below the clump. 
Other than the small number of bright main sequence stars,
we therefore employ stars of colors and magnitudes 
similar to those used in Galactic Bulge measurements
that use clump giants \citep{hamaa}.

\begin{figure}
 \centering
\includegraphics[width=7.8cm]{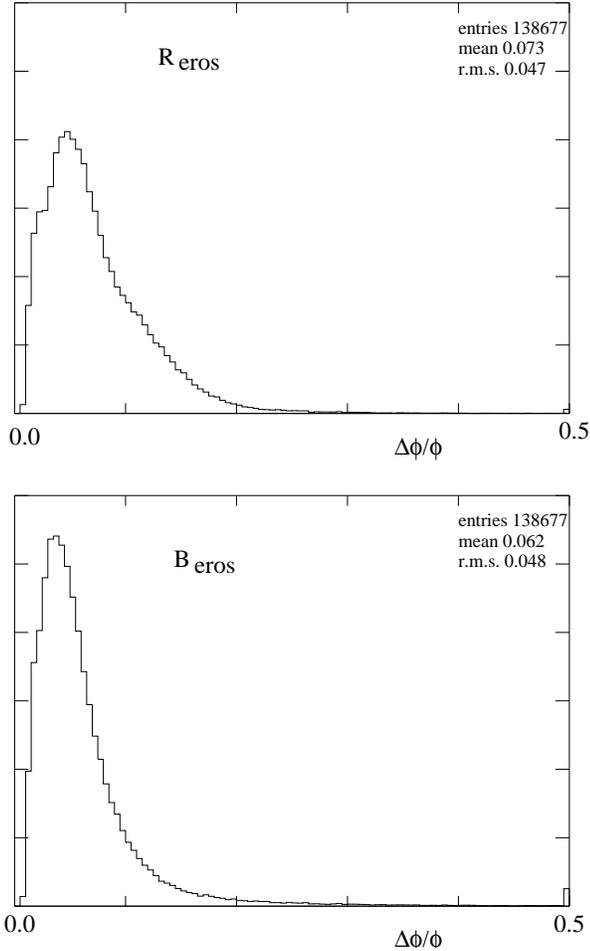}
 \caption{
The flux precision, $\Delta\phi/\phi$, for the
Bright-Star Sample  (2\% of the sample are in the
histograms).  
Each entry is the mean flux uncertainty for a star on its
light curve.
The top (bottom) histograms are for
the $\reros$  ($\beros$) bands. 
}    
\label{resbrightfig}
\end{figure}

It turns out that the cut (\ref{brightdefeq}) gives a rather uniform
photometric precision for the Bright-Star Sample.  Figure \ref{resbrightfig}
shows the distribution of precision with a mean in both bands
of $\sim7\%$.  The precision in dense fields hardly differs from
that in sparse fields and the precision in the LMC hardly differs
from that in the SMC.

The number of EROS-2 objects in the Bright-Star Sample
defined by (\ref{brightdefeq}) is
6.05 million in the
LMC and 0.90 million in the SMC.  
We must subtract from these numbers the expected
number of foreground stars in the Milky Way disk.
The Besan\c{c}on model of the Milky Way \citep{besancon}
predicts that 0.56 million (LMC) and
0.04 million (SMC) are foreground Milky Way stars,
consistent with the number of stars
we observe that are redder than Magellanic red giants.
The number of Bright-Sample stars to be used for optical
depth measurements must therefore be reduced by 9\% (4\%) for the LMC (SMC), 
giving
\begin{equation}
N_{objects}=5.49\times10^6 \hspace*{3mm}{\rm (LMC)} \;;
\end{equation}
\begin{equation}
N_{objects}=0.86\times10^6 \hspace*{3mm}{\rm (SMC)} \;.
\end{equation}

We now show how we evaluate the effect of blending on
the Bright-Star Sample.
For a given lens model (defined by the spatial,
velocity and mass distribution of lenses), the expected number of 
detected events is
\begin{equation}
N_{\rm ex}\;=\; 
\frac{2}{\pi}\,\frac{T_{\rm obs}}{\langle\te\rangle}\, \tau\,
\sum_{i=1}^{N_{\rm stars}} \langle \epsilon_i \rangle
\label{nexpectedeq}
\end{equation}
where $T_{\rm obs}$ is the observing period, $\langle\te\rangle$ is
the mean Einstein radius crossing time given by the model,
e.g. (\ref{smodelteeq}), and $\tau$ is
the optical depth.
The sum is over all observed stars and the ``efficiency'', 
$\langle \epsilon_i \rangle $,
is the expected ratio between the number of events passing 
the selection cuts of Section \ref{analsec} and
the number of events with $u<1$ occurring during the observation period.
The efficiency  
must be averaged over the $\te$ distribution given by the model.

EROS photometry is performed on
a set of ``objects'' found on  a reference image.  In crowded
fields, the correspondence between objects and stars is not
straightforward.  Generally speaking, most objects are dominated
by one star, i.e. by the brightest star in the object's
seeing disk.  However, if a fainter star in the seeing disk is
microlensed, EROS photometry will assign some of the extra flux to
the object with the remainder assigned to sky background.
It is thus convenient to rewrite (\ref{nexpectedeq}) as
\begin{equation}
N_{\rm ex}\;=\; 
\frac{2}{\pi}\,\frac{T_{\rm obs}}{\langle\te\rangle}\, \tau\,
\sum_{j=1}^{N_{\rm objects}}\, \sum_{i} 
\langle \epsilon_{ij} \rangle
\label{nexpectedeq2}
\end{equation}
where the second sum is over the stars $i$ near the object $j$.  
The efficiency, 
$\langle \epsilon_{ij} \rangle $,
is again the expected ratio (averaged over $\te$)
between the number of microlensings of 
star $i$ that pass selection cuts and
the number of microlensings of star $i$ 
with $u<1$ occurring during the observation period.
Since we do not know the stars in the seeing disk of each object,
we must estimate the sum statistically using knowledge of the 
luminosity function.
We will see that sum is dominated by the brightest star, $i=1$, with
contributions from the second brightest, $i=2$, of order 10\%.

If there were no blending, the
efficiency could be estimated by modifying a sample of light
curves with measured fluxes $F_0(t)$ by  
\begin{equation}
F(t) = F_{0}(t)\frac{u^2 + 2}{u\sqrt{u^2 + 4}} \, ,
\hspace*{5mm}        u(t)^2 = u_0^2 + [ (t - t_0) / t_E ]^2 \;, 
\label{noblendeq}
\end{equation}
where $u(t)$ is the distance of
the lens to the star's line-of-sight in units of the
Einstein radius, $u_0$ is this distance
at the time, $t_0$, of maximum amplification, and $\te$ is the
Einstein ring radius crossing time.
The modified lightcurves can then be subjected to the selection
criteria and the efficiency deduced by averaging over $\te$ and $u_0$.

For blended events only a fraction $\alpha$ of an object's
flux  is amplified.  If the fraction is independent of amplification,
the resulting light curve is
\begin{equation}
F(t) = F_{0}(t)\left[ (1-\alpha) + 
\alpha\frac{u^2 + 2}{u\sqrt{u^2 + 4}} \right] \, .
\label{blendeq}
\end{equation}
The efficiency to see such events now depends on $\alpha$:
$\epsilon_{ij}=\epsilon_{ij}(\alpha_i)$.  Stars
with small $\alpha_i$ require a small impact parameter $u_0$ to give
a sufficiently large reconstructed amplification.
Since our selection criteria require a reconstructed amplification
greater than 1.34 ($u_0=1$ for $\alpha=1$), 
the primary effect  of blending is to reduce the efficiency by 
a factor $u_i(\alpha_i)$,
the impact parameter needed for star $i$ to
produce a reconstructed amplification of 1.34.
The efficiency is reduced by this factor because of the flat
a priori $u$ distribution.  Additionally, the event appears
shorter by a factor $u_i$ further modifying the efficiency.

\begin{figure*}
\includegraphics[width=.9\textwidth]{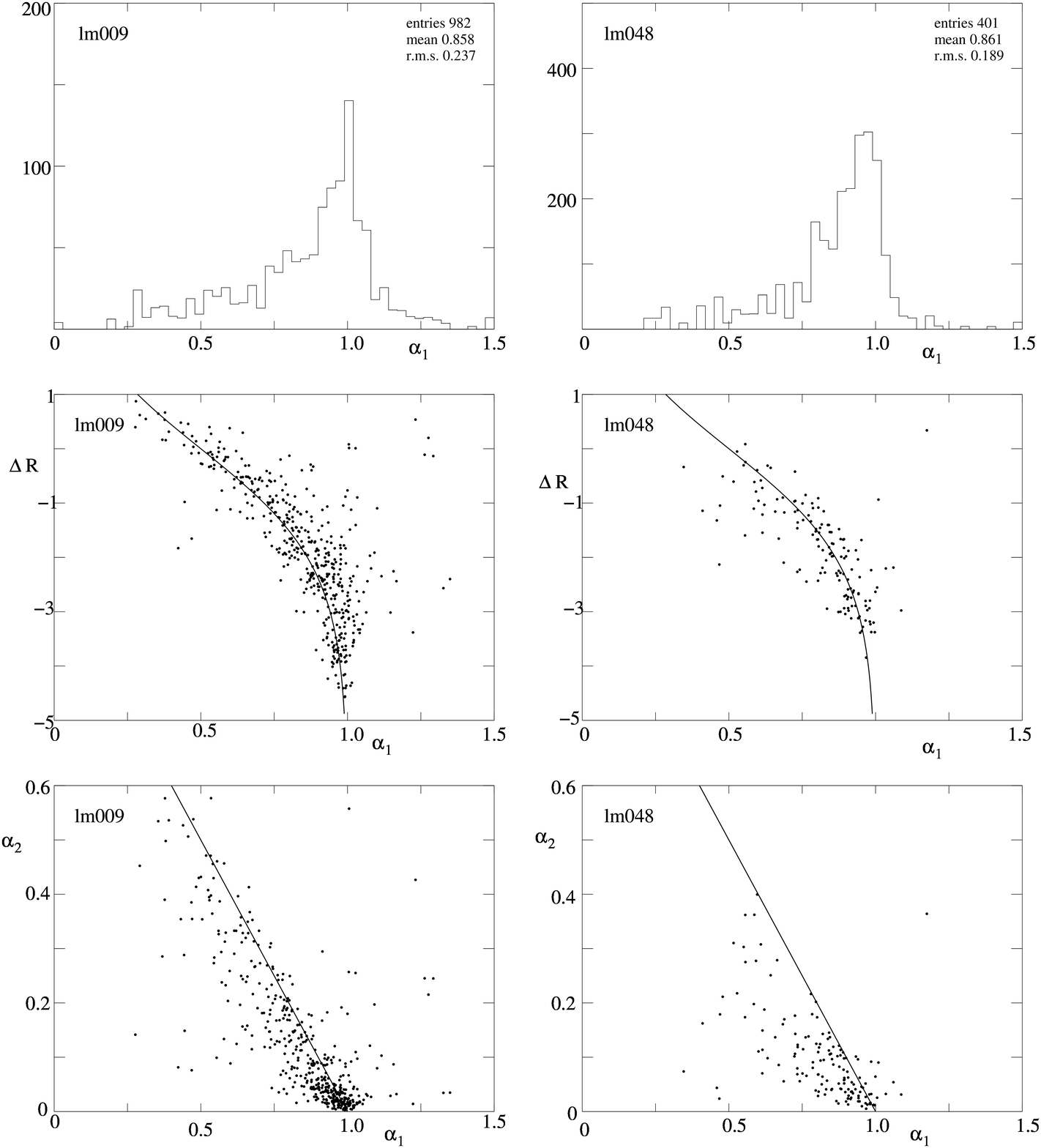}
\caption{
The effects of blending on artificial stars added to 
the dense field lm009 (left) and the sparse field lm048 (right).
The top two panels show the distribution of $\alpha_1$ using
artificial stars that do not fall under a brighter pre-existing
object.  Stars further than 2 arcsec from the nearest pre-existing
object are concentrated in the peak at $\alpha_1\sim 1$ while
stars nearer to pre-existing objects form the tail at 
 $\alpha_1 <1$.  The middle two panels show, for objects within
2 arcsec of pre-existing objects, the correlation between 
$\alpha_1$ and the difference in magnitude between the artificial
star and the pre-existing object.  Artificial objects with
$\alpha_1$ significantly below unity are  associated with
pre-existing objects of similar magnitude. 
The line shows the expected relation for two superimposed
objects: $\alpha_1=1./(1.+10^{0.4\Delta R})$.
The bottom two  panels
show the distribution of $(\alpha_1,\alpha_2)$ with $\alpha_2$
calculated from (\ref{alphadeq}) and the magnitude of the
pre-existing object.
The line shows the expected relation for two superimposed objects:
$\alpha_1 + \alpha_2 =1$.
}
\label{alphafig}
\end{figure*}

A simple  example is two superimposed stars of the
same magnitude and color.  In this case 
$\alpha_1=\alpha_2=0.5$ and $u_1=u_2\sim 0.7$.
Before efficiency corrections, the
event rate would be $0.7+0.7=1.4$ times the rate calculated ignoring blending.
If the two stars
have colors differing by 
$\Delta(B-R)_{eros}=0.7$, the requirement that $A_{max}>1.34$ in both
colors yields a rate that is increased by
a factor 1.3 over the rate for $\alpha=1$.

To statistically evaluate the distribution of the $\alpha_i$ for Bright-Sample
objects,
a spectrum of 8000 artificial stars was placed at 
random positions on real EROS images
(lm009, lm019, lm034, lm048)
to give artificial images (xm009, xm019, xm034, xm048).
These fields were chosen as representative of the crowding
variations over the LMC fields.
The stars were given fluxes according to randomly
chosen microlensing events  (one per artificial star).
Reference xm images were created using these images and photometry
performed in the same manner as for the normal (lm)
images.

After photometry, the light curves of xm objects falling near
artificial stars were studied to
find values of the $\alpha_i$.
In practice, one median value of $\alpha$ was calculated for each lightcurve
using points for which $(A-1)_{input}>0.5$, though no nonlinearity was
observed that made this a crucial point.
Of the
8000 artificial stars that 
were added to each of the four fields, most are
not usable for various (understandable) reasons:  star in a
masked region of the CCD, event during a period with no observations,
or the reconstructed magnitude
not satisfying the Bright Sample cuts.
For the field xm009, a total of 1123 lightcurves were usable for determining
$\alpha$.  Of these lightcurves, 982 concerned objects 
for which the artificial star
is the primary component, while in the remaining 141 objects the 
artificial star falls
underneath a preexisting bright object on the original image.
We use the
first type of object to determine the distribution of
$\alpha_1$, the value of $\alpha$ for the primary star associated
with each object.

Figure \ref{alphafig} (top) shows the distribution of $\alpha_1$ for
the dense field lm009 and the sparse field lm048.
The other two fields give similar distributions.
The distributions of $\alpha_1$ are characterized by 
peaks at $\alpha_1\sim1$ due
to artificial stars falling more than $\sim 2\,{\rm arcsec}$ from any
pre-existing lm object.  This happens for about half the artificial
stars in the densest field (lm009) so blending has little
effect on about half the bright stars in dense fields.  
Stars falling on pre-existing stars 
yield the tail at $\alpha_1<1$.  
This can be seen in the middle plots that show, for artificial
stars within $2\,{\rm arcsec}$ of pre-existing stars, the correlation 
between $\alpha_1$ and the magnitude difference between artificial
and pre-existing stars.  The distribution follows closely the
expected form $\alpha_1=1/(1+f_2/f_1)$  where $f_1$ is the flux of
the artificial star and $f_2$ is the flux of the preexisting star.
For example, when $\alpha_1\sim0.5$ the cloud of points passes
near $\Delta R\sim0$.

The mean value of 
$\alpha_1$ is $\sim0.86$ equivalent to a mean value of  
$u_1$ of 0.92.  This means 
that the (efficiency uncorrected)
rate of events with amplifications of the primary
component greater than 1.34 is
reduced by a factor 0.92 from the rate calculated assuming no
blending.

The loss of 8\% of the events is compensated by extra events due to stars
under the primary component.
Since $\alpha_1\sim1/(1+f_2/f_1)$, we can expect 
that $\alpha_2\sim1/(1+f_1/f_2)$ for superimposed objects.
However, the superposition is not perfect so $\alpha_2$ must
be a decreasing function of the separation between the 
object and star 2.
The artificial
events due to artificial stars falling under brighter pre-existing objects
are well described by 
\begin{equation}
\alpha_2(d) \;\sim\; \frac{f_2}{f_1+f_2}
\;exp[-(d/d_0)^2] \hspace*{10mm}d_0=1.9\,{\rm arcsec }
\label{alphadeq}
\end{equation}
where $f_1$   is the flux of the major component,  $f_2$
is the flux of the  minor
(amplified) component and
$d$ is the separation between the xm object and the minor component.
This formula can then be applied to the objects where the artificial
star is the major component since the flux and position 
of the minor component
(on the original image) is known.
In the bottom
panels of figure \ref{alphafig} we show 
$\alpha_l\,vs.\alpha_2$  for the
fields lm009 and lm048.  
The anticorellation between $\alpha_l$ and $\alpha_2$ means
that the loss of events due to $\alpha_1<1$ is partially 
compensated by events from the second star on an object by
object basis.

\begin{figure}
\includegraphics[width=7.5cm]{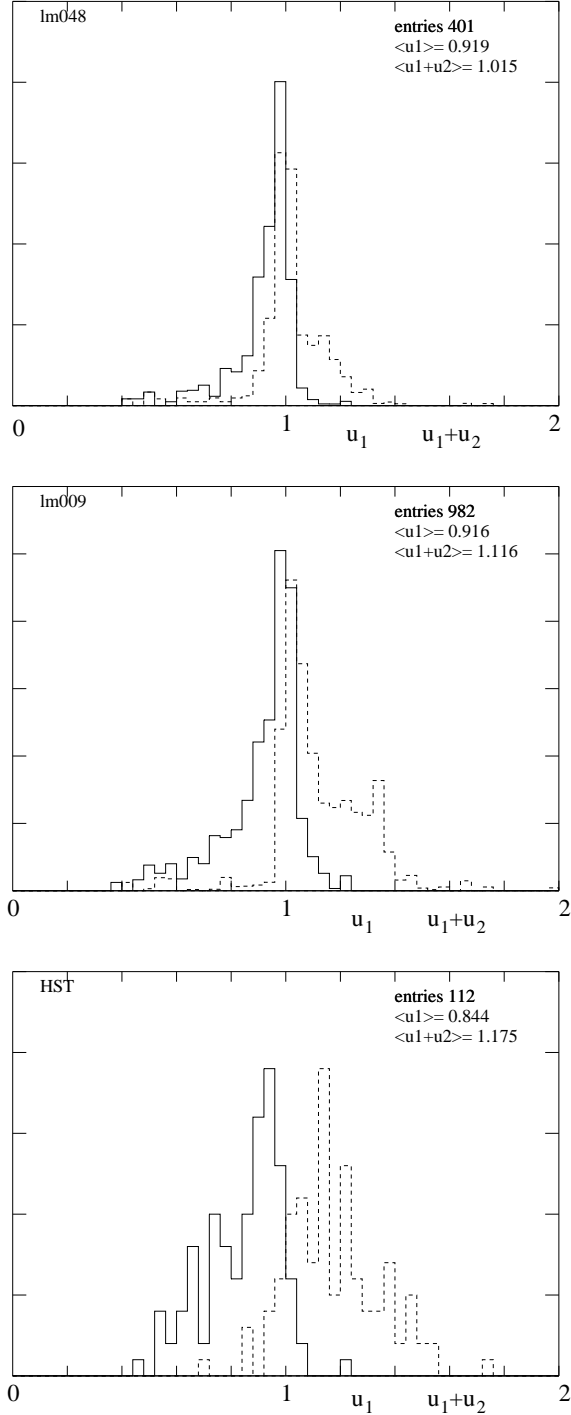}
\caption{
The distribution of $u_1$ (solid line) and
($u_1+u_2$) dashed line for the sparse EROS field lm048 (top panel)
the dense EROS field lm009 (middle panel) and the very dense HST
field in lm009, ccd3 (bottom panel) as described in the text.
The distributions of $u_1$ are all characterized by a peak
near $u_1\sim1$ and a tail at $u_1<1$.  With increasing
star density, the mean of $u_1+u_2$ increases because of
the increasing importance of secondary stars.
}
\label{usummcsynthfig}
\end{figure}

The compensation is seen in figure \ref{usummcsynthfig} histogramming
$u_1$ and $u_1+u_2$.  The sum  $u_1+u_2$
 gives the rate for event with amplifications
$>1.34$ compared to that calculated assuming $\alpha=1$.
In lm048, events due to the second star compensate
for the loss of events on the first star 
while in the dense field lm009 there is a 10\% overcompensation.
As we will see in Section \ref{effsec}, the efficiency to observe
the blended events is degraded compared to unblended events,
so the overall number of expected events is about 10\% less than
what one would calculate ignoring blending.

We have checked our calculations by studying public HST images of
a small part of the densest EROS region, that of  lm009 ccd 3.
A catalog of stars for this region was produced and images with
EROS seeing were fabricated.  As in the previous
analysis, images with some amplifications were compared with
reference images.  Figure \ref{usummcsynthfig} shows the distribution of $u_1$
for bright objects on the convoluted images.
Because of the very high density of stars, $u_1$ is somewhat
smaller than that calculated with the artificial
images: $\langle u_1\rangle=0.85$
as opposed to $\langle u_1\rangle=0.92$.  This lower value of 
$u_1$ is, as expected, compensated by higher values of $u_2$.
Figure \ref{usummcsynthfig} shows the 
histogram of $u_1+u_2$ indicating that the
event rate is raised by 15\% over the unblended rate.  Including
the loss of efficiency modifies this so that the event rate is
only 5\% higher than the unblended rate.

\section{Candidate selection from the Bright-Star Sample}
\label{analsec}

The present analysis aims at detecting luminosity excursions, due to  
microlensing, on otherwise constant light curves. The smallest reachable
microlensing timescale, $\te$,  is determined by the sampling of the fields; 
in practice, except for a few fields,
the detection efficiency is very low below 2-3~days.
The largest detectable timescale, about 800~days, is determined by the
reduction in the detection efficiency when the timescale becomes so
large that the baseline flux is not seen during the observing period.

The analysis is guided using the 
simulation of microlensing described in Section \ref{effsec},
which also serves to 
determine the efficiency of the selection procedure.

\begin{table}
\begin{center} \begin{tabular}{l r r  } \hline\hline
cut       & LMC & SMC  \\
\hline
\\
Bright-Star Sample             &              6052102  & 900809    \\              
Filter (C1, C2, C3)    &                   33876  &   5787    \\                
$LP_{N,pos,1}/LP_{N,pos,2} > 10$  (C4)  &   6778  &    847   \\
good $\chi^2_{\rm base}$  (C5)  &            2636  &    266   \\     
$t_0$ (C6)                          &       2065  &    195  \\     
$N_{peak}  > 4$  (C7)     &           1624  &    149 \\                  
significant $\Delta\chi^2$   (C8)      &              350  &     32  \\                      
good $\chi^2_{peak}$ (C9)  &                  152  &     17  \\                      
BlueBumper (C10)    &                         10  &      4  \\                    
SN ($S<0.3$) (C11)  &                          7  &      4  \\                     
$u_0 < 1$  (C12)         &                     1(SN)  &      1   \\                 
\\
\end{tabular}
\caption{
The number of light curves  surviving the cuts C1 to C12 as described
in the text.
}
\label{cuttable}
\end{center}
\end{table}

\begin{figure}
 \centering
\includegraphics[width=7.8cm]{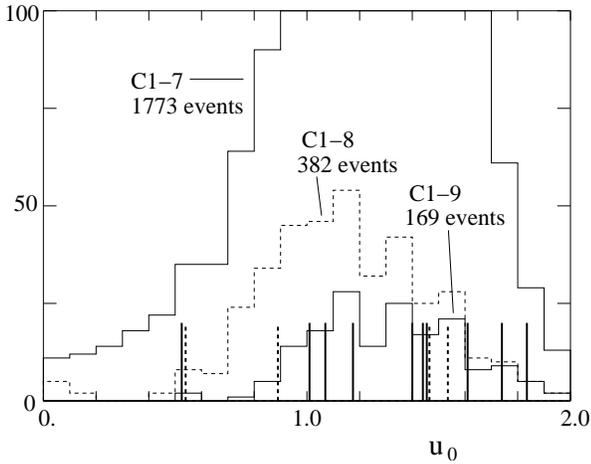}
 \caption{
The $u_0$ (mean of those for $\reros$ and $\beros$)
distribution for LMC and SMC light curves surviving cuts C1-C7, C1-C8,
and C1-C9.  The positions of the 14 light curves  
surviving cuts C1-C10 are shown with the 4 supernovae as dashed lines.
}    
\label{u0distfig}
\end{figure}

In the present analysis, luminosity excursions are defined with respect to
the baseline stellar flux; the first
task is thus to determine this baseline. To that end, we order the $N$ 
measured fluxes $\Phi_i$ of the light curve by increasing values of $\Phi$. 
For each of the $ N-1 $ middle values of the flux intervals 
$ \Phi_{i,mid} = 0.5 ( \Phi_i + \Phi_{i+1} ) $, we count the number 
$ N_{run,i} $ of runs, 
i.e. groups of consecutive points on the light curve that are on the same 
side of $ \Phi_{i,mid} $.\footnote{Or, equivalently, $ N_{run,i} $ is the 
number of times the time-ordered light curve crosses the constant 
flux line at $ \Phi_{i,mid} $, plus one.}
The baseline flux is defined as that value of 
$ \Phi_{i,mid} $ which maximizes $ N_{run,i} $. This way of defining
the baseline has a precision similar to that of a  simple average of the 
fluxes, but it proves much more robust to aberrant measurements, and 
less biased for most variable stars. We have checked on the simulated 
microlensing 
light curves that there is no visible bias for timescales $t_E$ shorter 
than $\sim 200\,\day$.   The bias is non-negligible above $600\,\day$, 
but this is taken into account in the efficiency calculation.

Next, we determine the point-to-point dispersion in the light curve 
$\sigma_{ptp}$, from the comparison of each measured flux with the linear 
interpolation of its two neighboring (in time) 
fluxes. This is done separately for
the 7 seasons of data taking, as we have observed a progressive degradation 
of the photometric scatter in the last 3 seasons.

Using the baseline flux and the photometric scatter, we then search for 
luminosity excursions defined as 
a group of consecutive points with fluxes
sufficiently far from
the baseline flux.
These should have at least 5 consecutive measurements
more than $ 1.5 \sigma_{ptp} $ from the baseline. 
In practice, to increase 
the detection efficiency for short duration phenomena, we allow small 
``holes'' within the excursion 
(series of points less than $ 1.5 \sigma_{ptp} $ from the baseline) 
provided that each hole contains
exactly one measurement. 
We call positive (negative) excursions those 
comprised of fluxes higher (lower) than the baseline.

The relative significance of each excursion is then estimated using the 
probability that it corresponds to a statistical fluctuation of a normal
law. We use the variable $ LP_N $, the co-logarithm of this probability
$$
 LP_N = - \sum_{i=1}^{i=N}
  \log\left(\frac{1}{2}\:{\rm Erfc}\left(\frac{x_{i}}{\sqrt{2}}\right)\right), 
$$
where $x_i$ is the deviation from the baseline
of the point taken at time $t_i$, in units of
its error $\sigma_i$, and $N$ is the number of points within the excursion.
The excursions 
are ranked by decreasing values of $ LP_N $ (decreasing significance).

We now describe the 12 selection criteria, $C1$-$C12$, used to
select microlensing candidates.
$C1$ requires at least one significant positive excursion:
\begin{displaymath}
C1: \hspace*{5mm} LP_{N,pos,1} > 20  \;.
\end{displaymath} 
The largest negative excursion should be much smaller than the largest 
positive one: 
\begin{displaymath}
C2: \hspace*{5mm}LP_{N,pos,1} / LP_{N,neg,1} > 10  \;.
\end{displaymath}
 There should 
be less than 10 excursions in total:
\begin{displaymath}
C3: \hspace*{5mm} N_{\rm excursions}<10 \; .
\end{displaymath} 

Cuts $C1$, $C2$ and $C3$ are applied independently in the two passbands and a star
is retained for further analysis only if it passes the three cuts
in both bands. 
There are 33876 LMC light curves and 5787 SMC light curves
that survive.
The progression of the number of surviving light curves
with the subsequent cuts is shown in Table \ref{cuttable}.

All light curves passing $C1-3$ are  fitted with a simple microlensing 
curve (\ref{noblendeq}), 
independently in the two passbands. The results of the fits, i.e. the fitted 
parameters $t_0$, $u_0$ and $t_E$ and the values of $\chi^2$, are used in 
cuts C5-C10.  

Cuts C4 and C5 eliminate light curves with significant variability
outside the main positive excursion.
We first require that the second positive excursion be much less significant 
than the first one:
\begin{displaymath} 
C4:\hspace*{5mm}  LP_{N,pos,1} / LP_{N,pos,2} > 10 \;.
\end{displaymath} 
To discriminate against low amplitude variations 
(either real or due to photometric
problems) we require that the normalized $\chi^2_{base}$ 
for a time independent 
light curve outside the main excursion (points with $u>2$)
be sufficiently good:
\begin{displaymath} 
C5:\hspace*{5mm}  \frac{ \chi^2_{base} - N_{dof,base} }
{  \sqrt{2 N_{dof,base} }}\; 
<\; 20 \;.
\end{displaymath}
This cut may seem
loose as it would correspond to cuts at 20 standard deviations, 
for a normal law. However, 
the errors we use to compute $ \chi^2_{norm}$  
are only estimates of the actual errors. For isolated stars, the errors 
are overestimated. The values of the cuts have thus been tuned using the 
simulated light curves, that have the same uncertainties as the original 
data; the two cuts reject about 10\% of the simulated microlensing light 
curves.

Cuts C6 and C7  require that the main fluctuation occur at a time when
the light curve is sufficiently well sampled.
In order to discriminate against stellar variations with  very asymmetric peaks
we thus require that the time $t_0$ be 
within the  observing
interval:
\begin{displaymath}
C6:\hspace*{5mm}  T_{beg} + \te \;<\;t_0 \;<\; T_{end} - \te \;,
\end{displaymath} 
where $T_{beg}$ 
and $T_{end}$ are the first and last date of EROS-2 observations. 
We also require that there be more than 
4 flux measurements within the main fluctuation ($u<2$)
\begin{displaymath}
C7:\hspace*{5mm}N_{peak}>4  \;.
\end{displaymath}

There are 1624 LMC light curves and 149 SMC light curves that
survive cuts C1-C7.  Their distribution of $u_0$ and those of
light curves satisfying subsequent cuts  are
shown in Figure \ref{u0distfig}.  It is seen that most light curves
show only low amplitude variations with $u_0>0.6$.

We next require that the
microlensing fit to the light curves in both passbands be significantly 
better than the fit of a constant flux:
\begin{displaymath}
C8:\hspace*{5mm}
\frac{\chi^2_{ct} - \chi^2_{ml}}{\chi^2_{ml} / N_{dof}}  
\frac{1}{\sqrt{2 N_{dof,peak}}} \,>40 \;,
\end{displaymath}
where $\chi^2_{ct}$ and $\chi^2_{ml}$ are the chi-squared values for the
constant fit and the simple microlensing fits, and $N_{dof}$ and 
$N_{dof,peak}$ are the number of measured points in the full light curve
and in the peak. 
The event is required to pass this cut 
in both passbands.
This cut eliminates light curves that have low amplifications, poor sampling,
and/or poor photometric errors.

We then require a reasonably good fit to the microlensing curve
within the peak:
\begin{displaymath}
C9:\hspace*{5mm} 
\frac{ \chi^2_{peak} - N_{dof,peak} }
{  \sqrt{2 N_{dof,peak} }}\;
<\; 10 \;.
\end{displaymath}

There are 152 LMC and 17 SMC light curves that pass cuts C1-C9.
As can be seen in Figure \ref{u0distfig}, almost all of these light curves have
$u_0>0.8$.  They are mostly bright
main-sequence stars that display 
low-amplitude variations that are difficult to distinguish
from microlensing, though they often show $\sim20\%$ more
variation in the red than in the blue.
They were first mentioned in \citet{alc97a} and are commonly
called ``blue bumpers''.

We identify blue bumpers  first from their position in the color-magnitude
diagram and their low, chromatic amplification.
For stars with $(B-R)_{\rm eros}<0.2$ and
$\reros<18.6$ (LMC) or $\reros<18.9$ (SMC), we require
\begin{displaymath}
C10:\hspace*{5mm}min(A_R,A_B)>1.6 \hspace*{5mm}{\rm and}
\hspace*{5mm}
\frac{A_R-1}{A_B-1}<1.2 \;,
\end{displaymath}
where $A_R$ and $A_B$ are the maximum amplifications observed
in the two bands.  These two amplifications are taken to
be the fitted amplifications at the time of the maximum {\it observed}
amplification.  This value is used since the fitted amplification
at $t=t_0$ may significantly overestimate a blue bumper amplitude
if the peak region is not well sampled.

There are 10 LMC light curves and 4 SMC light curves that
pass cuts C1-C10.  Four of these light curves are most likely
supernov\ae\ (SN) exploding in 
galaxies far behind the Magellanic Clouds. 
In the analysis of the Full Sample of EROS-2 
Magellanic light curves \citep{tis04}, 31
such supernovae were found.
All SN in our sample have fitted timescales in the range 25-50~days, 
asymmetric light curves with a faster rise time, and larger variations in 
the bluer passband. For a fraction of them, about 20\%, the host galaxy is visible, 
which makes them indisputable SN. In order to identify the remaining SN, 
we have devised a fitting function with a time asymmetry parameter $S$; 
the function reduces to simple microlensing for $S = 0$. The fitting 
function is obtained from simple microlensing (\ref{noblendeq})
by replacing the Einstein timescale $t_E$ by the ``supernova'' 
varying timescale 
\begin{equation} 
         t_S = t_E \: \left[ 1 + S \: \arctan 
     \left( \frac{t - t_0}{t_E}\right) \right] \: .
\end{equation}
For positive (negative) values of $S$, the rise time is faster (slower) 
than the decrease time.  When fitting this function to the remaining light 
curves, we expect to find $S > 0$ for the SN.
Based on the study of the supernovae found in the Full Sample analysis
and on simulated microlensing events, 
we choose to eliminate
light curves with blue-band asymmetries by requiring
\begin{displaymath}
C11: \hspace*{5mm} S_b \;<\; 0.30 \;. 
\end{displaymath}
This cut eliminates 3 light curves 
from the Bright-Star Sample, leaving 11 light curves, mostly with very
small amplitudes, $u_0>1$.
Since the $u_0$ distribution for microlensing events is flat, the
events with $u_0>1$ in Figure \ref{u0distfig} must be mostly background.
We therefore require
\begin{displaymath}
C12:\hspace*{5mm}u_0<1.0 \;.
\end{displaymath}
This leaves two light curves.  One of them is superimposed on a clear
background galaxy.  It is therefore most likely a supernova and
we eliminate this event leaving one event to be discussed in the
next section.

\section{Candidate sample and verifications}\label{sec:candidates}
\label{candsec}

\begin{figure*}
\includegraphics[width=.95\textwidth]{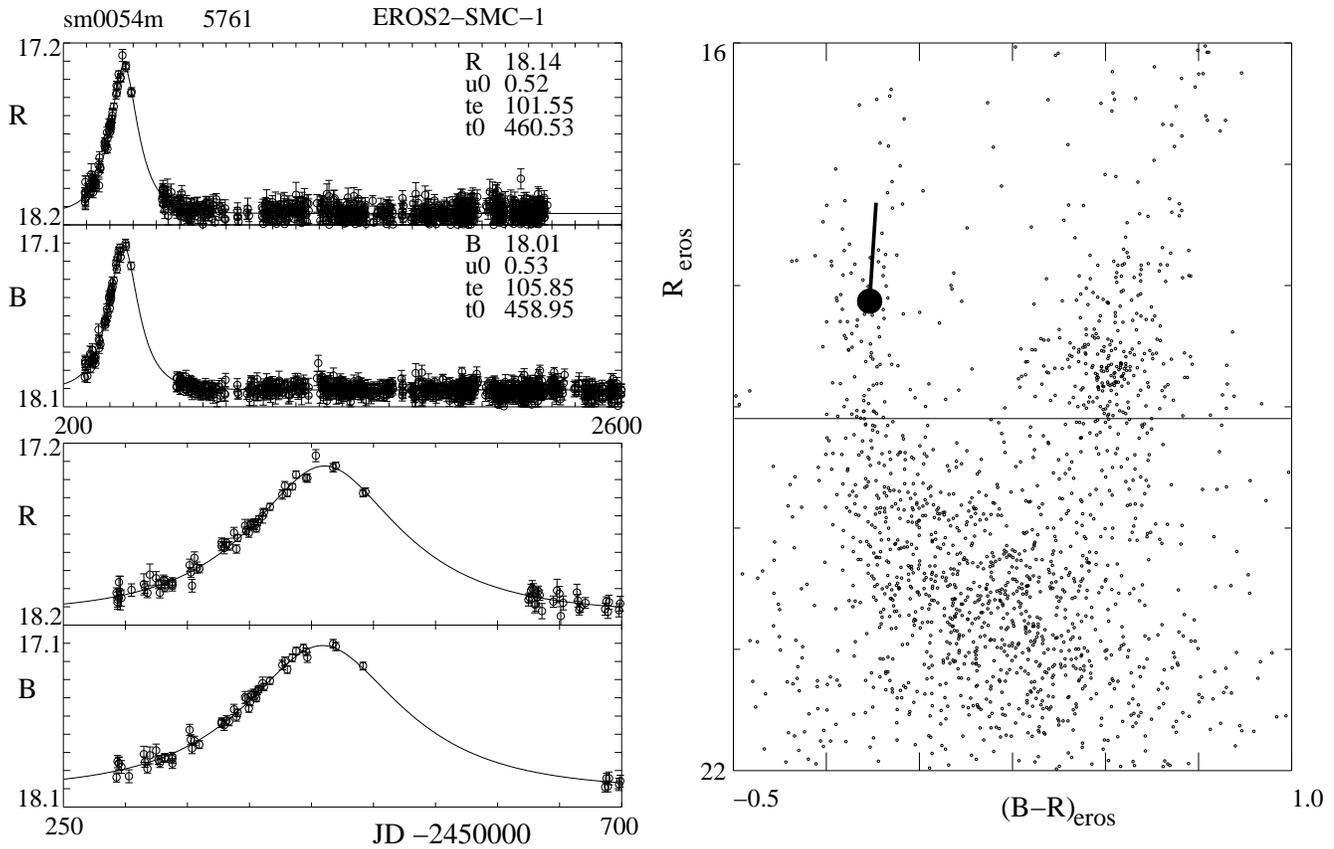}
 \caption{
The light curves of EROS-2 microlensing candidate EROS2-SMC-1 
(star sm005-4m-5761).
Also shown is the color-magnitude diagram of the star's CCD-quadrant
and the excursion of the event.}
\label{smc1fig}
\end{figure*}

\begin{figure*}
\includegraphics[width=.95\textwidth]{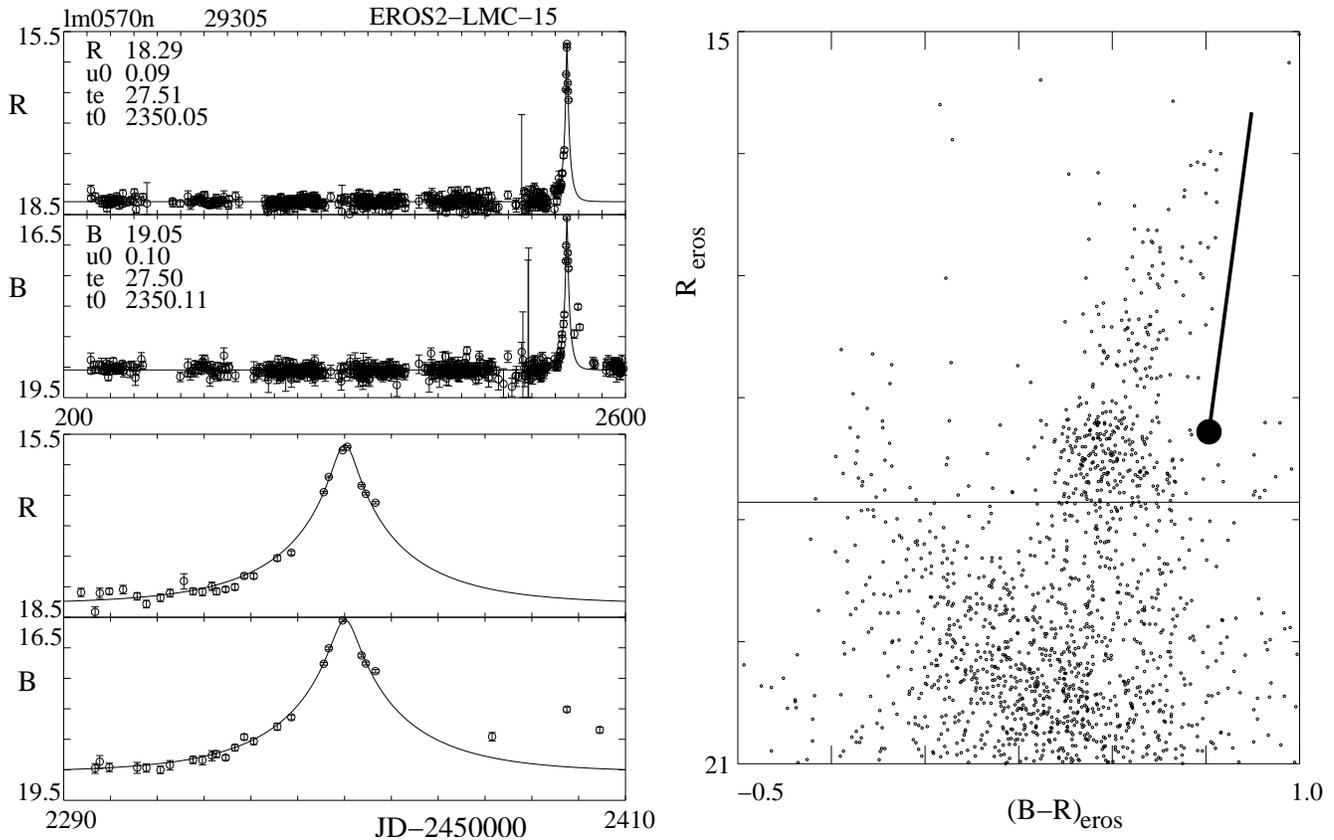}
 \caption{
The light curves of EROS-2 star lm057-0n-29305 (r.a.$=79.9488\,\deg$, dec.$=-70.7741\,\deg$). 
Also shown is the color-magnitude diagram of the star's CCD-quadrant
and the excursion of the event.}
  \label{lm057fig}
\end{figure*}

After the cuts described in the previous two sections have been applied,
only one candidate microlensing event remains in the  Bright-Star Sample, 
EROS2-SMC-1.
Its light curve is
shown in fig. \ref{smc1fig} and its characteristics are
given on the first entry in Table \ref{pateventtable}. 
It has been known since 1997 \citep{smc1macho,smc1} and is
identical to candidate MACHO-97-SMC-1. It has one of the two
longest timescales ($103\,\day$) and the highest luminosity of all 
published microlensing 
candidates reported toward the Magellanic Clouds. 
The star is separated by $1.6\,{\rm arcsec}$ from another star that
is $1.2\,{\rm mag}$ fainter 
\citep{uda97} causing $\sim30\%$ blending in EROS-2 images.
Including the blend in the light curve fit increases 
$\te$ to $125\,\day$.\footnote{When blending is taken into account, this
event has the longest $\te$ of any published Magellanic event.}
The star displays a 4-5\% variability with a period 
$P = 5.12$~d \citep{smc1}. For further details on this event, see 
\citet{smc1}, \citet{smc2, EROSMC} and \citet{parsmc}.

\begin{table*}
\begin{center} \begin{tabular}{l l l l l l l l l l  } \hline\hline
candidate & EROS-2 star & r.a. (deg) & dec. (deg) & $\reros$  & $(B-R)_{\rm eros}$ & $t_0$ & $\te$ & $u_0$ & $\chi^2/dof$ \\
\hline
\\
EROS2-SMC-1  & sm005-4m-5761  &   15.0233 &  -72.2507 &   18.14 &   -0.13 &   459.0 &   105.8 &     0.53 &   972.4/873 \\
             &                &            &            &         &         &   460.5 &   101.6 &     0.52 &   862.4/748 \\
EROS2-SMC-5  & sm006-6k-19475 &   12.6625 &  -72.6983 &   17.83 &    0.34 &  1414.8 &    29.5 &     0.90 &   485.1/875 \\
             &                &            &            &         &         &  1432.0 &    15.1 &     1.12 &   560.0/573 \\
EROS2-LMC-8  & lm055-7m-23303 &   76.6396 &  -71.6624 &   20.34 &    1.16 &  1594.1 &    15.3 &     0.01 &   344.0/572 \\
             &                &            &            &         &         &  1594.1 &     8.0 &     0.03 &   540.9/510 \\
EROS2-LMC-9  & lm042-1l-2622  &   75.3717 &  -65.0486 &   20.93 &    0.23 &  2204.1 &    57.8 &     0.36 &   640.8/369 \\
             &                &            &            &         &         &  2215.3 &    52.7 &     0.30 &   292.1/228 \\
EROS2-LMC-10 & lm045-5n-26323 &   80.9925 &  -65.8733 &   20.55 &    0.81 &  1600.3 &    37.1 &     0.33 &   580.0/505 \\
             &                &            &            &         &         &  1602.2 &    35.9 &     0.51 &   649.1/447 \\
EROS2-LMC-11 & lm061-4m-15782 &   88.4392 &  -71.2536 &   20.55 &    0.55 &   451.9 &    51.3 &     0.36 &   456.5/444 \\
             &                &            &            &         &         &   446.6 &    46.1 &     0.37 &   458.7/394 \\
EROS2-LMC-12 & lm085-6l-14234 &   89.8750 &  -74.5676 &   20.19 &    0.34 &  2664.0 &    49.6 &     0.27 &   580.4/391 \\
\\
\end{tabular}
\caption{
Events found in the analysis of the full sample of 
EROS-2 stars \citep{tis04}.
Only event EROS2-SMC-1 
is in the Bright-Star Sample and passes the cuts C1-C12 and therefore
is used for limits on the optical depth.
All fits assume zero blending and no intrinsic variability.
The two values of the fitted parameters $t_0$, $\te$ and $u_0$ are
for $\beros$ (first line) and $\reros$ (second line).  (There are no
$\reros$ points during the event for EROS2-LMC-12.)
The time of maximum, $t_0$ is given as JD-2450000.
}
\label{pateventtable}
\end{center}
\end{table*}

As a check on the cuts $C1-12$ leading to EROS2-SMC-1 and to
search for nonstandard microlensing events
(e.g. those due to binary lenses),
a large number of light curves were visually scanned.
Among them were 
all events satisfying $C1$-$C8$ and $u_0<1$,
all events satisfying $C1$-$C7$ and $u_0<0.5$,
all events satisfying $C1$-$C10$,
and all events satisfying $C1$-$C12$ but with $R_{max}$ 
increased by $0.2\,{\rm mag}$.
Only one interesting event was found, EROS2-LMC-15
shown in Figure \ref{lm057fig}.
The beginning of the event is very similar to a standard microlensing
event but subsequent points are too high.\footnote{OGLE-3 data confirm
the high flux of the post-peak points (A. Udalski, private communication).}
It was rejected  by the  $\chi^2_{peak}$ cut 
($C9$)  but it may, in fact, be a lensing
event due to a binary lens.
Note that its position in the color-magnitude 
diagram suggests that the star is either
not in the LMC or behind a foreground Milky Way star.

An important verification of the analysis of the Bright-Star Sample
presented here 
comes from an analysis of the
Full Sample of 
Magellanic light curves described in \citet{tis04}.
Because faint stars were analyzed, the cuts,
given in Appendix \ref{fullselectionsec}, were generally slightly
stricter than those described in Section \ref{analsec}.
However, the analysis did not require observation of the
event in both colors.  It therefore provides an important
check on our analysis.  The candidates found in this
analysis are listed in Table \ref{pateventtable} and their
light curves shown in Appendix \ref{fullselectionsec}.
Two candidates EROS2-SMC-1 and 5 are in  the Bright-Star Sample.
Candidate 5 does not pass the cuts presented here because
it fails $C12$.  We also note that it shows non-microlensing-like
variations in the light curve of the MACHO 
collaboration\footnote{http://wwwmacho.mcmaster.ca/Data/MachoData.html.
There is no information at this site on the EROS-2 LMC candidates 
of Table \ref{pateventtable} since
none are in MACHO fields.}.

Two other independent analyses were performed on
the SMC data, restricted to the first five years of data.
In both cases, the only event found in the Bright-Star Sample was
EROS2-SMC-1. The first analysis, reported in 
\citet{EROSMC}, was based on a larger set of stars (5.2 million); 
the number of analysis cuts common to it and the present analysis 
is small; the computer programs were written independently. The 
second analysis followed a complete reprocessing using a new 
implementation of differential photometry developed by \citet{triton}.
The technique would allow us to find events not on cataloged stars.

We attempted to check our efficiency for finding
microlensing events by considering the events published
by the MACHO collaboration.
The 13-17 events \citep{macho57} used by them to measure
the optical depth toward the LMC are listed in
Table \ref{machoeventtable}.  
Only 2 of the 17 stars (MACHO-LMC-18 and 25)
are bright enough to be in our Bright-Star Sample.
Three of the 17 events occurred after the beginning of EROS-2 
operations but none of these three events occurred on stars
in our Bright-Star Sample.  
One of the three (MACHO-LMC-15) was on a star too dim
to be seen by EROS-2.\footnote{
The $1.27\,{\rm m}$ telescope of
the MACHO collaboration allowed them to use fainter stars than
EROS-2 with its $1\,{\rm m}$ telescope.
}  
The two other events 
(MACHO-LMC-14 and 20) are seen in the
EROS-2 images and give microlensing parameters  compatible with
those measured by MACHO (see Table \ref{machoeventtable}).

MACHO-LMC-14 was selected in the MACHO A analysis.
In EROS-2, it was located in a defective zone of 200~x 2000 pixels on 
CCD 7 of our blue camera, near the edge of the mosaic.
The corresponding star was not cataloged in this band, and the star
failed the requirement to be observed in both passbands. 
Consequently, it does not appear in Table \ref{pateventtable}.

MACHO-LMC-20 was selected in the MACHO B analysis.
In EROS-2, it is located 35 arcsec
from a very bright Galactic star (about $V = 10$ vs. 21 for the 
candidate). 
In this analysis, stars too near bright stars were
eliminated so this candidate does not appear in Table \ref{pateventtable}.

\begin{table*}
\begin{center} \begin{tabular}{l l l l l l l l   } \hline\hline
MACHO candidate       & EROS-2 star & $\reros$ & $(B-R)_{\rm eros}$ & $t_0$ & $\te$ & $A_{max}$    & comments \\
   & & & & & & &      \\
\hline
\\
  MACHO-LMC-1      & lm019-6k-14879 &      18.97 &       0.61 &     -942.9 &      17.20 &       7.15 &          \\
   MACHO-LMC-4     & lm057-0l-27696 &      19.85 &       0.24 &    -353.1 &      22.70 &       2.92 &          \\
   MACHO-LMC-5     & lm057-0k-27469 &      19.97 &       1.52 &    -976.1 &      37.80 &      47.28 &          \\
   MACHO-LMC-6     & too faint     &            &            &     -802.9 &      45.80 &       2.43 &          \\
   MACHO-LMC-7     & lm010-2k-21383 &      20.47 &       0.52 &    -536.5 &      51.45 &       5.91 &          \\
   MACHO-LMC-8     & too faint &                 &            &     -612.2 &      33.20 &       2.19 &        \\
   MACHO-LMC-9 B-binary & lm001-0l-6864 &      20.92 &       0.83 &  -400.1 &      89.60 &       1.95 &       \\
  MACHO-LMC-13     & lm020-6m-19733 &      21.03 &      -0.07 &    134.0 &      50.05 &       2.36 &          \\
  MACHO-LMC-14     & lm002-7n-24616 &      18.84 &            &    391.3 &      50.05 (38.2) &       3.37 (2.77) &          \\
  MACHO-LMC-15     & too faint     &            &            &    472.4&      18.40 &       2.83 &          \\
  MACHO-LMC-18     & lm060-2n-16057 &      18.35 &       0.47 &    -217.2&      37.10 &       1.54 & Bright Sample   \\
  MACHO-LMC-20 B   & lm012-5n-23157 &      20.09 &       1.03 &    774.7 &      36.35 (23.2) &       2.95 (3.1) &          \\
  MACHO-LMC-21     & dead zone      &            &            &   -786.8 &      46.60 &       5.64 &          \\
  MACHO-LMC-22 B   & dead zone      &            &            &    -43.3 &     114.65 &       2.70 &          \\
  MACHO-LMC-23     & lm055-3n-4994 &      19.82 &       0.37 &    -237.7 &      42.60 &       2.41 & variable (Fig. \ref{macho23fig}) \\
  MACHO-LMC-25     & lm017-2m-2056 &      17.95 &            &    -265.7 &      42.60 &       1.50 & Bright Sample   \\
  MACHO-LMC-27 B   & lm010-0n-23830 &      19.00 &       0.14 &   -485.9&      25.25 &       1.45 &          \\
\\
\end{tabular}
\caption{
The 17 microlensing candidates of the MACHO collaboration \citep{macho57}.
Candidates 20, 22 and 27 (marked B)
are low signal-to-noise candidates  not satisfying the 
MACHO ``A'' requirements.
Candidate 9 is due to a binary lens and does not satisfy their A requirements. 
The values of $\te$ and maximum amplification
$A_{max}$ are from the MACHO fit assuming no blending, and $t_0$ is given
as JD-2450000.
Candidates 18 and 25 are each within $1\,{\rm arcsec}$ of an object of similar
magnitude that is resolved by the MACHO analysis but blended in the EROS
analysis.
Candidates 14 and 20 occurred during EROS-2 observations, and the
EROS-2 values of $\te$ and $A_{\rm max}$ are shown in parentheses
(averages of $\reros$ and $\beros$ measurements).
}
\label{machoeventtable}
\end{center}
\end{table*}

\section{The status of published Magellanic microlensing candidates}
\label{formersec}

In this section, we review and update the status of the published
Magellanic microlensing candidates of the EROS and MACHO collaborations.

EROS has, in the past, used 11 candidates to place upper limits on the
optical depth toward the LMC and to measure
the depth toward the SMC.  Six of the candidates are in the Bright-Star Sample
considered here.
The candidates and their present
status are given in Table \ref{eroseventtable}.

\begin{table*}
\begin{center} \begin{tabular}{l l l l l l } \hline\hline
Candidate & EROS-2 star     & $\reros$ & $(B-R)_{\rm eros}$ & original ref.  & status  \\
\\
\hline
\\
EROS1-LMC-1     & lm058-2k-21915    &   18.75 &    0.34 & \citet{aub93} & $2^{nd}$ variation  \citep{tis04} (Figure \ref{eros1-1fig})       \\
EROS1-LMC-2     & lm043-6m- 9377 B-S  &   19.32 &   -0.04 & \citet{aub93} & $2^{nd}$ variation    \citep{las00}       \\
EROS2-LMC-3 & lm034-6l-20493    &   20.90 &    0.61 & \citet{las00} &  $2^{nd}$ variation   \citep{tis04}       \\
EROS2-LMC-4 & lm018-6n-23236    &   19.10 &    1.87 & \citet{las00} &   $2^{nd}$ variation  \citep{mil00}       \\
EROS2-LMC-5 & lm015-3n-22431 B-S  &   19.17 &    0.14 & \citet{mil00} &  Supernova  \citep{tis04}                 \\
EROS2-LMC-6 & lm067-5m-14700    &   21.01 &    0.63 & \citet{mil00} &  Supernova  \citep{tis04}                 \\
EROS2-LMC-7 & lm070-3n-23389    &   21.00 &    0.76 & \citet{mil00} &  Supernova  \citep{tis04}                 \\
EROS2-SMC-1 & sm005-4m-5761 B-S  &   18.13 &   -0.13 & Palanque-D. et al.(1998)&  Fig.  \ref{smc1fig}                      \\
EROS2-SMC-2 & sm001-6l-13221 B-S  &   19.56 &    0.44 & \citet{EROSMC}&  long period variable \citep{tis04}       \\
EROS2-SMC-3 & sm001-6n-16904 B-S  &   19.31 &    0.59 & \citet{EROSMC}&  long period variable \citep{tis04}       \\
EROS2-SMC-4 & sm002-7m-21331 B-S  &   19.48 &    0.32 & \citet{EROSMC}&  long period variable \citep{tis04}       \\
\end{tabular}
\caption{
The 11  events of the EROS collaboration used in the past to set upper
limits on the microlensing optical depth toward
the LMC and to measure the depth toward the 
SMC.  
Candidates marked ``B-S'' in column 2 occur on stars in the EROS-2 
Bright-Star Sample.
All candidates except EROS2-SMC-1 
have been eliminated as variable stars or as supernovae.
}
\label{eroseventtable}
\end{center}
\end{table*}

Of the candidates
in Table \ref{eroseventtable} only EROS2-SMC-1 remains as a candidate
microlensing event.  The others have been eliminated 
either because  continued observations of the same stars show further 
variability on the light curves, or  because
improved photometry, in some cases complemented by spectroscopy, 
lead to re-interpreting the candidates as variable stars.
Before the analysis 
presented here, the following EROS candidates were eliminated:
EROS1-LMC-2, presented in \citet{aub93} and which 
displayed a new variation 8 years later \citep{las00}; 
and candidate EROS2-LMC-4, presented in \citet{las00} 
and eliminated in \citet{mil00}.  

Before this analysis started, there were 5 surviving LMC microlensing 
candidates from EROS, one from EROS-1 (number EROS1-LMC-1) and four from 
EROS-2 (numbered EROS2-LMC-3, 5, 6 and 7).

\begin{figure*}
\includegraphics[width=.9\textwidth]{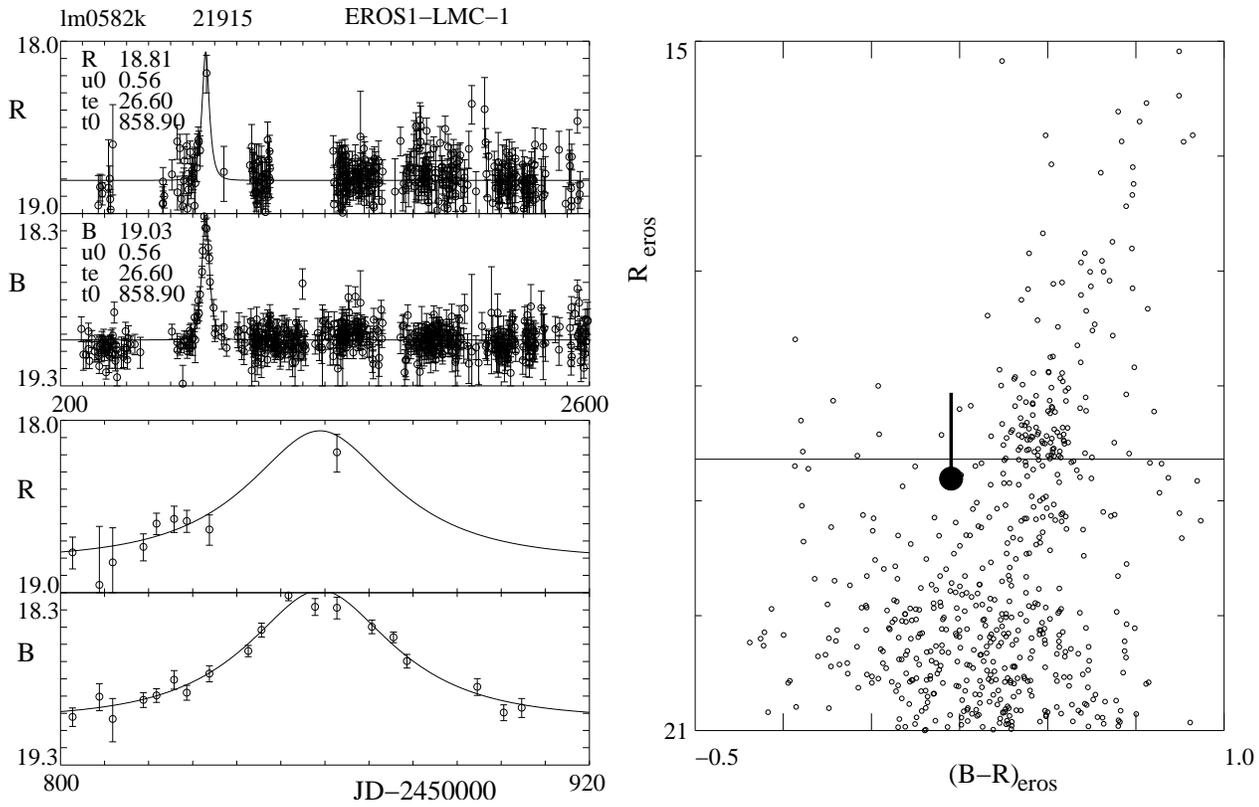}
 \caption{
The EROS-2 light curve of EROS-1 microlensing candidate EROS1-LMC-1.  
The curve shows
a second variation, 6.3 years after the variation observed in EROS-1. 
Also shown is the color-magnitude diagram of the star's CCD-quadrant
and the excursion of the event.}
\label{eros1-1fig}
\end{figure*}

The EROS-1 candidate EROS1-LMC-1 displayed a new variation 
in the EROS-2 data (Figure \ref{eros1-1fig})
in 1998, 6.3 years after the first one, of similar amplitude (a factor 
two) and timescale (27~days). This second variation is well fitted by a 
microlensing light curve.  Because they are separated in time by more 
than 80 Einstein timescales, the probability that these two bumps 
correspond to the microlensing of a double source star is lower than 
half a percent, even in the favorable case of the two stars being of equal
luminosity. 
This candidate was thus rejected. Let us recall that 
it was already known to be a Be star \citep{beau95} and was thus 
already suspected of being variable \citep{Pac96}. Candidate EROS2-LMC-3 also 
displayed new variations between 1999 and 2002, of a more irregular 
nature, and was thus rejected.

The light curves of candidates EROS2-LMC-5, 6 and  7 have been improved, 
due to the 
better template images in the present analysis. The reduced photometric 
scatter, compared to \citet{mil00}, made apparent an asymmetry in rise 
and fall times; the asymmetry test (C11) allowed us to 
identify and reject them as supernov\ae. 
EROS2-LMC-5 is identical to MACHO-LMC-26, 
which had been rejected by \citet{macho57} for the same reason.

The conclusion is that {\it none} of the former EROS LMC microlensing 
candidates are still considered valid. Four displayed further variability, 
and three were identified as SN thanks to improved photometry.

There have been four EROS-2 SMC candidates, 
EROS2-SMC-1  discussed in Section \ref{candsec}, and
EROS2-SMC-2, 3 and 4 
presented in \citet{EROSMC}. 
Candidates 2, 3 and 4  were described as doubtful candidates, 
as all three display very long timescale variations, and look
more like irregular variable stars. Another analysis of the same SMC
data using differential photometry \citep{triton}
allowed us to reduce the photometric errors and substantiate this 
interpretation.  The additional two years of data
since \citet{EROSMC} have  confirmed this, as further
irregular variations have been observed.   
They are not selected in the present analysis, and are now considered 
as definitively rejected.

\begin{figure*}
\includegraphics[width=.9\textwidth]{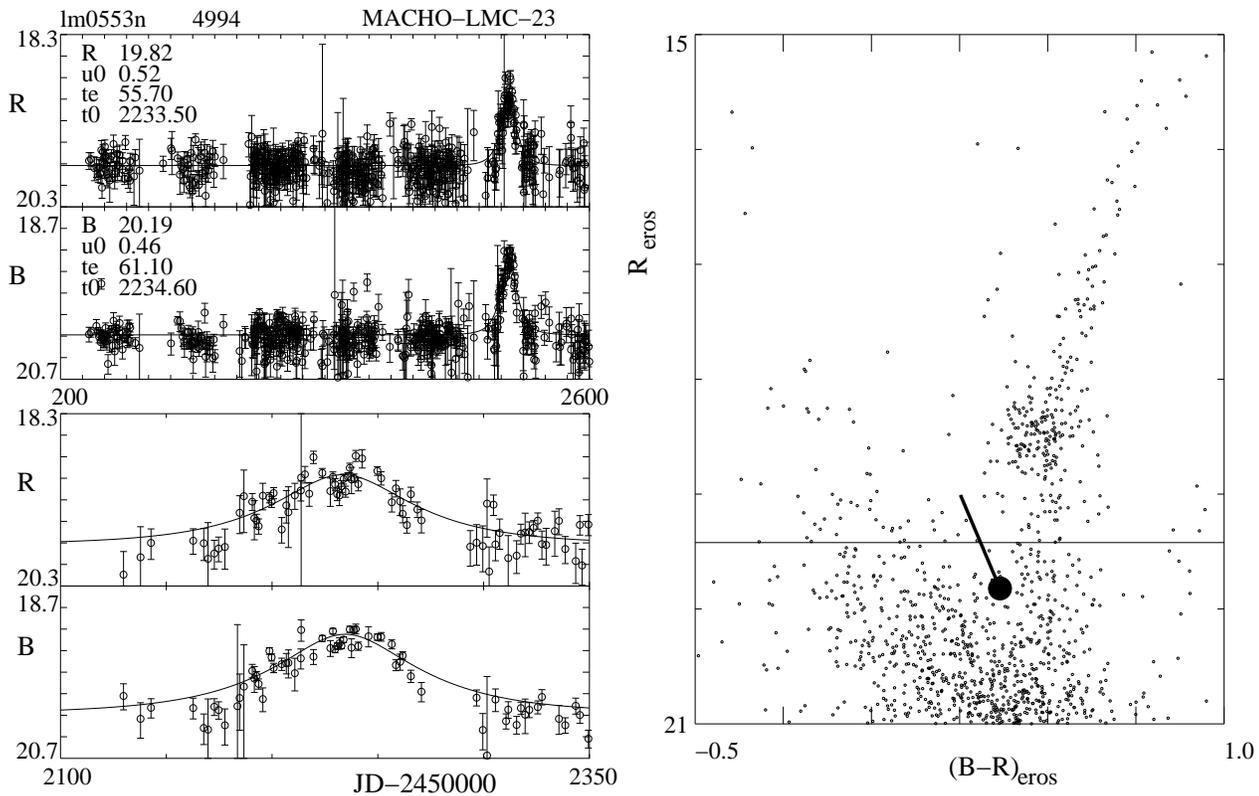}
 \caption{
The EROS-2 light curve of MACHO microlensing candidate MACHO-LMC-23.
The curve shows a second variation, 6.8 years after the variation
seen by MACHO.
Also shown is the color-magnitude diagram of the star's CCD-quadrant
and the excursion of the event.}
\label{macho23fig}
\end{figure*}

We have also searched for further variations in the
17 stars used by the MACHO collaboration to measure
the optical depth toward the LMC (Table \ref{machoeventtable}).  
One of the candidates, MACHO-LMC-23, showed a further variation
in the EROS-2 data \citep{tis04}, 
6.8 years after its first variation in the MACHO data.
Its EROS-2 lightcurve is shown in Figure \ref{macho23fig}.  
As such, we can eliminate it as a microlensing
candidate.  
In spite of this, both the variation shown in Figure \ref{macho23fig} and
the original variation in the MACHO data are quite achromatic,
indicating that achromaticity  is not a fool-proof criterion for 
selecting microlensing events.
We note however that 
\citet{bennett23} argued that, 
even without considering its
further variation seen in the EROS-2 data, 
the form of its light curve made MACHO-LMC-23 a weak microlensing candidate.

The MACHO collaboration has also reported candidate events
found by their alert system.  The most notable is MACHO99-LMC-2 
that was studied by \citet{macho99lmc2}.  This impressive microlensing
candidate was on a star too faint to be in the EROS Full Sample.

Besides the 17 LMC events of the MACHO collaboration, they have
reported two candidates in the SMC, though they have reported no
systematic search for SMC events.
The first MACHO SMC candidate \citep{smc1macho} 
is identical to EROS2-SMC-1 (Section \ref{candsec}). The second
candidate, MACHO-98-SMC-1, concerned a star too faint
to be included in our Full Sample of SMC stars.
The event
was detected by the alert 
system of the MACHO group in May 1998 and a probable caustic 
crossing due to a double lens was announced soon after.
The source star was monitored by most microlensing groups,
including EROS-2. The second caustic crossing was fully measured,
which allowed the  determination of the relative proper motion between
the lens and source, allowing
one to conclude that the event was due
to a lens in the SMC \citep{erossmcdouble,smcdouble}.   
\footnote{
Note that there exists a third, unpublished, microlensing 
candidate toward the SMC, OGLE-2005-SMC-001 
(http://bulge.astro.princeton.edu/$\sim$ogle/). 
Its long timescale ( $> 150\,\day$), large amplification ( $> 12$)
and bright source ($I = 18.2$) offer the prospect of completely 
resolving the microlensing parameters degeneracy through 
measurements of lightcurve
deformations, e.g. of that due to parallax.
}

\section{The detection efficiency}\label{sec:effi}
\label{effsec}

To measure the optical depth from the detected events, or limits
on this quantity, we first need to evaluate the detection efficiency 
as a function of the time scale $t_E$.  
This was determined by using 
Monte Carlo simulated light curves~: we superimpose artificial
microlensing events on a representative sample of light curves, 
corresponding to 2\% of the Bright-Star Sample  from each of the 98 
monitored fields.  The light curve for a simple microlensing event
(i.e. point-source point-lens zero-blending) is described by 
three parameters of (\ref{noblendeq}):  
date of maximum amplification $t_0$, impact parameter 
$u_0$ and time scale $t_E$. 
Blended events have the additional parameter $\alpha$ in
(\ref{blendeq}).
The microlensing 
parameters are chosen at random: $t_0$ follows a flat distribution over
our 2500~days observing period, JD 2,450,242 till 2,452,742; $u_0$ is 
picked randomly between 0 and 1.2; and $t_E$ is chosen at random from a 
distribution flat in $\ln (t_E)$, between 1 and 1000~days. Each star in 
the 2\% sample is actually used thrice in the simulation, once per decade 
in $t_E$. The simulation takes into account the relative variation of 
photometric errors.  

Simulated light curves were then fed into the analysis chain to find 
the fraction that are recovered by our detection algorithm. The detection 
efficiency in a given $t_E$ bin, $\epsilon (t_E)$,
is then given by the ratio between the  
number of events passing all selection criteria in this bin and the
number of microlensing events generated in the same bin with $u_0<1$.

\begin{figure}
 \centering
\includegraphics[width=7.8cm]{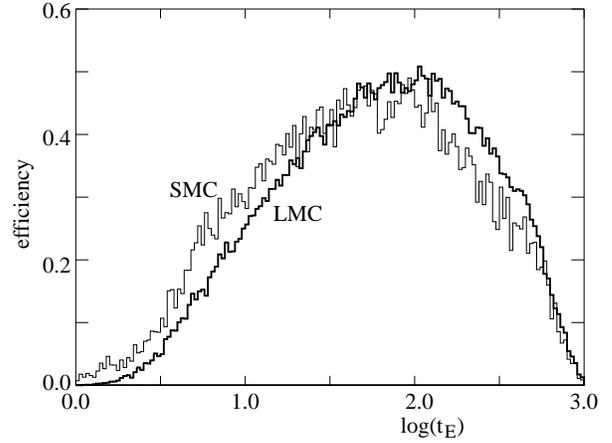}
 \caption{
The detection efficiency for unblended ($\alpha=1$) microlensing 
light curves (\ref{noblendeq})
as a function of $\te$ for LMC (bold line) and
SMC (light line) fields.  The efficiency applies to a $2500\,\day$
total observing period.
The efficiency used for calculation of the optical depth
is the efficiency shown here, multiplied by 0.9 to take
into account lensing by binary lenses.
}    
  \label{efffig}
\end{figure}

\begin{figure}
 \centering
\includegraphics[width=7.8cm]{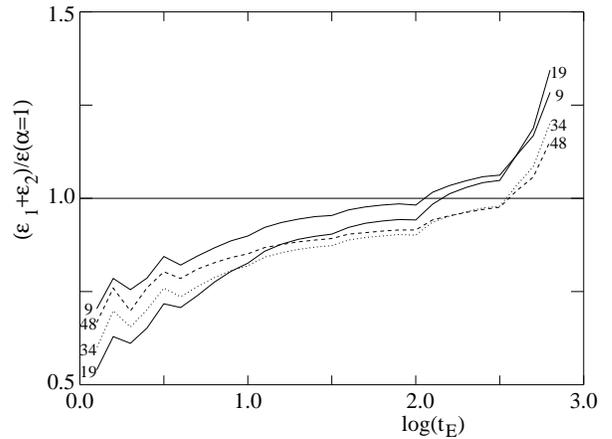}
 \caption{
The ratio between the summed efficiency, 
$\epsilon_1(\alpha_1)+\epsilon_2(\alpha_2)$ and the unblended efficiency
$\epsilon(\alpha=1)$ for pairs $(\alpha_1,\alpha_2)$ taken from the
artificial star analysis of Section \ref{brightsec}.  
The ratio is a function of $\te$.
The four curves
correspond to the four studied fields, lm009, 019, 034 and 048.
}    
  \label{effblendfig}
\end{figure}

\begin{figure}
 \centering
\includegraphics[width=7.8cm]{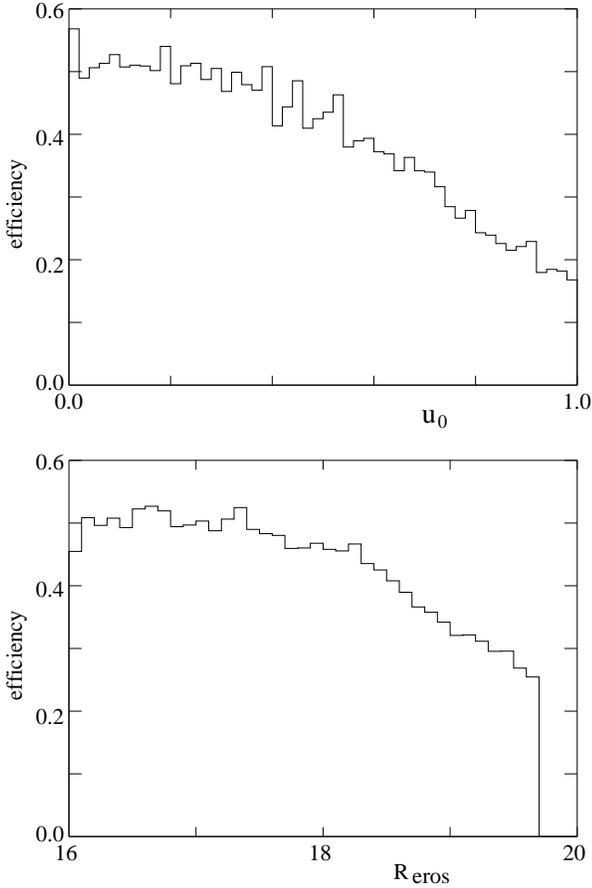}
\caption{
The detection efficiency for simple microlensing 
light curves (\ref{noblendeq}) as a function of $u_0$ (top)
and of $\reros$ (bottom) for LMC and SMC 
events in the range $10\,\day<\te<200\,\day$.
}    
  \label{effu0fig}
\end{figure}

Figure \ref{efffig} shows the LMC and SMC efficiencies for 
unblended events ($\alpha=1$) as a
function of $\te$.  
For the range of $\te$ of interest for this work, the efficiency
varies from $\sim0.25$ at $\te=10\,\day$ to $\sim0.45$ at $\te=200\,\day$.

Blended events have an efficiency that is reduced because an impact
parameter smaller than unity is necessary to produce an amplification
of 1.34.  
The efficiency is also modified because the time scale is reduced.
To sufficiently good approximation we find
\begin{equation}
\epsilon_{ij}(\alpha_i,\te) \;\sim\; \frac{u_{i}(2.18,\alpha_i)}{0.5}
\, \epsilon(\alpha =1,\,u_{i}(1,\alpha_i)\te)
\end{equation}
where $u_i(A_{max},\alpha_i)$ is the impact parameter necessary
to produce a maximum amplification $A_{max}$.

To evaluate the summed detection efficiency for a realistic distribution
of $\alpha_1$ and $\alpha_2$ we used the pairs from the artificial
images of Section \ref{brightsec}, Figure \ref{alphafig}. 
Figure \ref{effblendfig} shows the ratio between the calculated
sum (for the four studied fields) and the unblended efficiency.
For $\te=40\,\day$ the efficiency is reduced by a factor ranging
from 0.90 in the sparse fields lm048 and lm034 to 0.92 and 0.97 
in the denser fields lm019 and lm009.
The brightest star, $i=1$ accounts for 95\% of the rate in the sparse
fields and 88\% in the dense fields.

Figure \ref{effu0fig} shows the efficiency for unblended events
as a function of $u_0$ and of $\reros$ 
for events with $10\,\day<\te<200\,\day$.
The efficiency for the Bright-Star Sample 
has a much weaker dependence on  $u_0$ and  $\reros$
than that for the Full Sample.  For the Full Sample
the efficiency  falls rapidly with increasing $u_0$
and $\reros$ \citep{tis04}.

The efficiencies in Figures \ref{efffig} and \ref{effu0fig} are for
the detection of  microlensing events due to simple lenses.
Events due to binary lenses 
with caustic crossings are discriminated
against, mostly by $C9$.  
Of the 17 LMC events of the MACHO collaboration \citep{macho57} 
only 1 event, MACHO-LMC-9,
is of this type and would not pass our selection criteria.
This event corresponds to $\sim10\%$ of their optical depth.
We note also that toward the Galactic Bulge $\sim10\%$ of the 
observed microlensing events are
due to binary lenses \citep{UDA00}.  
To compensate for
this loss of efficiency, we conservatively reduce the efficiency 
of Figure \ref{efffig} by a factor $0.9$
when calculating limits on the optical depth.

\section{Limits on the abundance of machos}\label{sec:limits}
\label{limitssec}

The microlensing optical depth, $\tau$, is defined as the probability that
any given star, at a given time, is amplified by at least 1.34,
i.e. with an impact parameter $u < 1 $. 
From a set of $N_{\rm ev}$ events, $\tau$ can be
estimated from 
\begin{equation}
\tau \;=\; \frac{\pi N_{\rm ev}}{2N_{\rm stars}T_{\rm obs}}
\left\langle \frac{\te}{\epsilon}
\right\rangle \;,
\label{tauestimateeq}
\end{equation}
where $\langle\te/\epsilon \rangle$ is the mean $\te$ divided by 
efficiency for the observed events.

\begin{figure}
 \centering
\includegraphics[width=7.8cm]{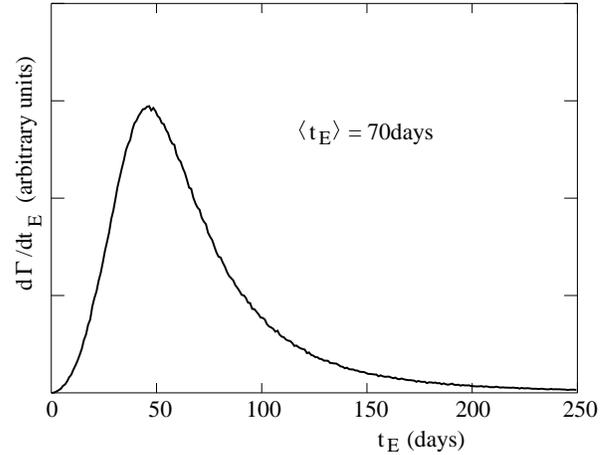}
\caption{
The $\te$ distribution $d\Gamma/d\te$ expected for 
$1M_\odot$ lenses in a spherically
symmetric isothermal halo with core radius $5\,\kpc$, i.e. the
S model used by the MACHO collaboration \citep{macho57}. 
}    
\label{tetheoryfig}
\end{figure}

In the LMC we have found no events so we can only
give an upper limit on $\tau$ by calculating
the expected number of events as a function of $\tau$
as given by (\ref{nexpectedeq2}).
For this analysis we use the $\te$ distribution of the S model
\citep{macho57}
shown in Figure \ref{tetheoryfig}.  It relates
$\langle\te\rangle$ to the macho mass (assumed unique):
$\langle\te\rangle=70\,\day\sqrt{M/M_\odot}$.
Limits using other halo models or macho mass distributions 
can  be found to often good approximation
by simply scaling (\ref{nexpectedeq2}) 
with $\epsilon(\langle\te\rangle)/\langle\te\rangle$.

\begin{figure}
 \centering
\includegraphics[width=7.8cm]{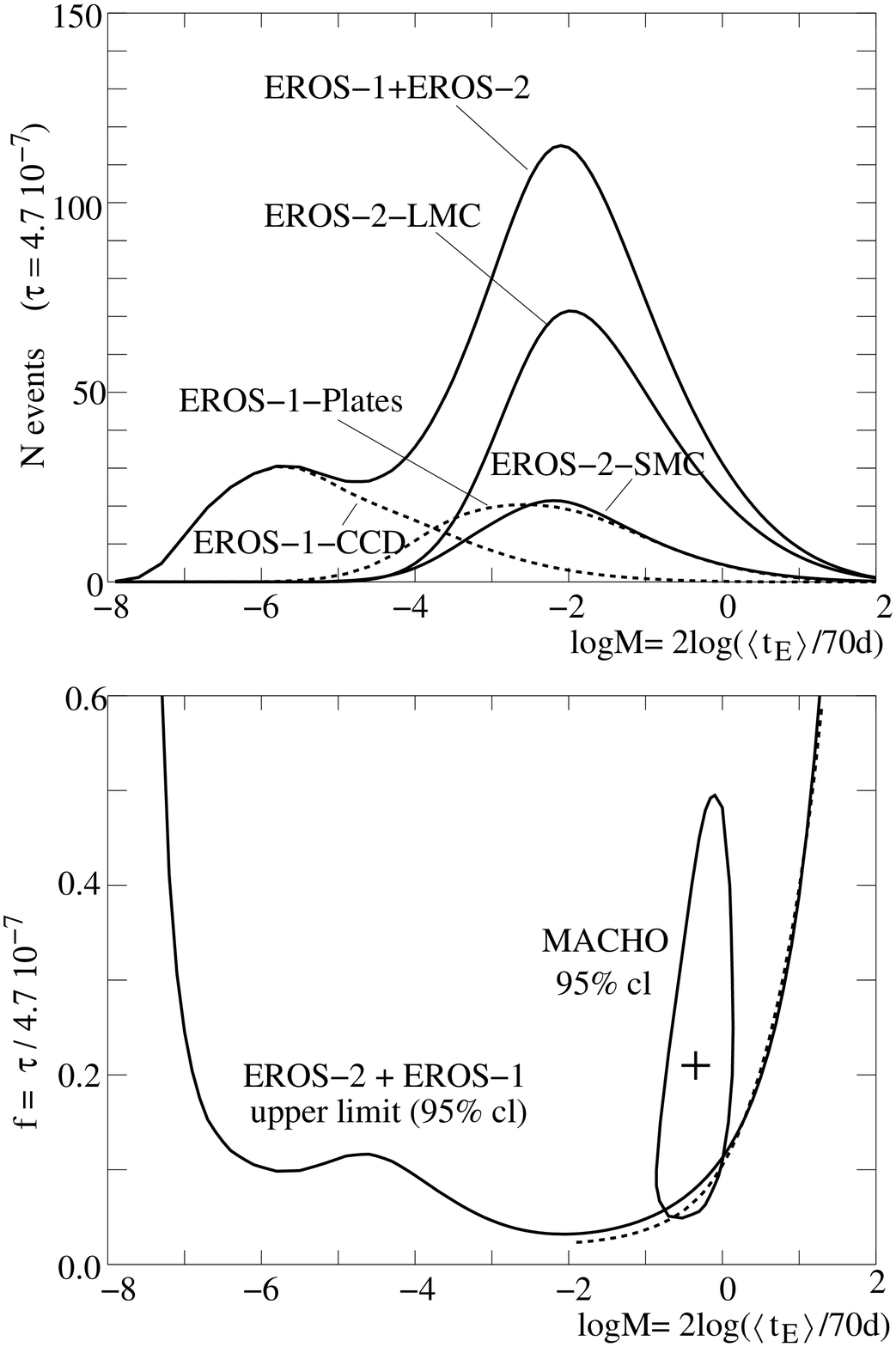}
 \caption{
The top panel shows the numbers of expected events as a function of 
macho mass $M$ for the S model of \citet{macho57}.
The expectations for EROS-2-LMC, SMC (this work) are shown along with
those of EROS-1 \citep{renault97} with contributions
from the photographic plate program  \citep{erosplates} and
CCD program \citep{renault98}.
The number of events for EROS-2-SMC supposes $\tau_{smc}=1.4\tau_{lmc}$.
In the lower panel 
the solid line shows the EROS 95\% CL upper
limit on $f=\tau_{\rm lmc}/4.7\times10^{-7}$ 
based on no observed events in the EROS-2 LMC data and the
EROS-1 data.
The dashed line shows the EROS upper
limit on $\tau_{\rm lmc}$ based 
on one observed SMC event in
all EROS-2 and EROS-1 data assuming
$\tau_{\rm smc-halo}=1.4\tau_{\rm lmc}$.  
The MACHO 95\% CL. curve is taken from Figure 12 (A, no  lmc halo)
of \citet{macho57}.
}   
\label{nevtsfig}
\end{figure}

The expected number of LMC events  for $\tau_{lmc}=4.7\times10^{-7}$
as a function of 
lens mass, $M$, is shown in Figure \ref{nevtsfig}a.  
For $M=0.4M_\odot$, we have $\langle\te\rangle=44\,\day$,
$\langle\epsilon\rangle=0.35$, $N_{star}=5.5\times10^6$ 
and $T_{obs}=2500\,\day$,
giving 32 LMC events for EROS-2.  We add 7 LMC events for EROS-1 to give
a total of 39 expected events for $\tau_{lmc}=4.7\times10^{-7}$.

For no observed events ($N<3.0$, 95\% CL),
the 95\% CL upper limit on the optical depth
is
\begin{equation}
\frac{\tau}{4.7\times10^{-7}} \;<\; \frac{3}{N_{ex}(4.7\times10^{-7})}
\label{genlimiteq}
\end{equation}
For 39 expected events,
The  upper limit is then
$\tau_{lmc}<0.36\times10^{-7}$.
The limit on 
$\tau_{\rm lmc}$ as a function of $M$ is shown in
Figure \ref{nevtsfig}b.  In the $\te$ range favored by 
the MACHO collaboration, we find
\begin{equation}
\tau_{\rm lmc} \;<\;0.36 \times 10^{-7} \,
\times \left[1 + \log (M/0.4M_\odot)  \right]
\hspace*{5mm}95\% CL \;,
\label{eroslimit}
\end{equation}
i.e.
\begin{equation}
 f \;<\; 0.077 \,
\times \left[1 + \log (M/0.4M_\odot)  \right]
\hspace*{5mm}95\% CL \;,
\label{erosflimit}
\end{equation}
where $f\equiv\tau_{\rm lmc}/4.7\times10^{-7}$ is the halo mass
fraction within the framework of the S model.
This limit on the optical depth is significantly
below the value for the central region of the LMC
measured by the MACHO collaboration 
\citep{macho57}, 
$\tau_{\rm lmc}/10^{-7}=1.2^{+0.4}_{-0.3}(stat.)\pm0.36(sys.)$ 
and the revised value of \citet{bennetttau}, 
$\tau_{\rm lmc}/10^{-7}=1.0\pm0.3$.
The \citet{macho57} optical depth used for
the entire LMC  predicts that EROS
would see $\sim9$ LMC events whereas none are seen.

For the SMC, the one observed event corresponds to an optical depth of 
$1.7\times 10^{-7}$
($N_{\rm star}=0.86\times10^6$).  
Taking into account
only Poisson statistics on one event, 
$0.05<N_{\rm obs}<4.74$ (90\% CL) this gives
\begin{equation}
0.085\times10^{-7}\;<\;\tau_{\rm smc} < 8.0\times 10^{-7} 
\hspace*{5mm}90\% CL\;.
\end{equation}
This is consistent with the expectations of lensing by objects in the 
SMC itself,
$\tau_{\rm smc}\sim0.4\times10^{-7}$ \citep{graffsmc}.
The value of $\te=125\,\day$ is also consistent with expectations for 
self-lensing $\langle\te\rangle\sim 100\,\day$
for a mean lens mass of $0.35M_{\odot}$.

We also note  that 
the self-lensing interpretation is favored from the absence of an 
indication of parallax in the light curve
\citep{parsmc}.

We can combine the LMC data and the SMC data to give 
a limit on the
halo contribution to the optical depth by supposing that the SMC optical
depth is the sum of a halo contribution, 
$\tau_{\rm smc-halo}=\alpha\tau_{\rm lmc}$  
($\alpha\sim1.4$)
and a self-lensing contribution $\tau_{\rm sl}$.
(We conservatively ignore contributions from LMC self-lensing
and from lensing by stars in the disk of the Milky Way.)
For one observed SMC event with $\te=125\,\day$ and zero 
observed LMC events, the likelihood function
is 
\begin{displaymath}
L(\tau_{\rm lmc},\tau_{\rm sl}) \,\propto\, 
\left[
\alpha\tau_{\rm lmc}\Gamma^{\prime}_{\rm h}(\te) 
+\tau_{\rm sl}\Gamma^{\prime}_{\rm sl}(\te)
\right]
\exp\left[-\,N(\tau_{\rm lmc},\tau_{\rm sl})\right]
\end{displaymath}
where $N(\tau_{\rm lmc},\tau_{\rm sl})$ is the total number of expected 
events (LMC and SMC) 
as a function of the two optical depths as calculated with 
equation (\ref{nexpectedeq}).
The function $\Gamma^{\prime}_{\rm h}(\te)$ is the distribution 
(normalized to unit integral)
expected for halo lenses of mass $M$ (Figure \ref{tetheoryfig}) and 
$\Gamma^{\prime}_{\rm sl}(\te)$ is
the expected distribution for SMC self-lensing taken from \citet{graffsmc}.
We assume the SMC self-lensing optical depth is that calculated by
\citet{graffsmc} though the results are not sensitive to this assumption.
For macho masses less than $1M_{\odot}$, 
the likelihood function is maximized for 
$\tau_{\rm lmc}=0$ because there are no
LMC events in spite of the greater number of LMC source stars.
For $M<0.1M_\odot$ the limit on the halo contribution 
approaches that one would calculate for no candidates in
either the LMC or the SMC
because
the observed $\te$ of $125\,\day$ is too long for a halo event.  
The calculated upper limit is shown as the dashed line  
in Figure \ref{nevtsfig}b.  In the mass range favored by the MACHO
collaboration, the limit is slightly lower than that using only the
LMC data.
The combined limit would be somewhat stronger if we assumed 
an oblate halo ($\alpha<1.4$)
and somewhat weaker if we assumed a prolate halo ($\alpha>1.4$).
Constraints on the shape of the Milky Way halo were recently  summarized
by \citet{fellhauer} who  argued that the
observed bifurcation of the Sagittarius Stream can be explained
if the halo is close to spherical.

A possible systematic error in our result could come from our assumption
that the optical depth due to binary lenses is small, 10\% of the total.
An alternative strategy would have been to relax the cuts so as to include
the event shown in Figure \ref{lm057fig}.
We have chosen not to do this because the light curve itself 
is not sufficiently
well sampled to establish the nature of the event (other than that it is not
a simple microlensing event) and also because of its anomalous position in the
color-magnitude diagram.  We note also that the optical depth
associated with the event, $\tau=0.7\times10^{-8}$, is 
a factor $\sim4$ below the
upper limit (\ref{eroslimit}).

Another important question concerns the influence on our results
of the  Bright-Sample magnitude cut.  Since the cut was not established
before the event search, it is natural to ask if the position of the
cut was chosen to give a strong limit.  In fact, elimination of the
cut would not change significantly the conclusions of this paper.
Four additional events 
(EROS2-LMC-8, 9, 10 and 11 from Table \ref{pateventtable})
were found with the analysis described in Appendix \ref{fullselectionsec}.
The baselines of the source stars of these events are
on average  0.9 magnitude below the Bright Sample cut.  An additional
event (EROS2-LMC-12) with no $\reros$ data during the variation was found.
The values of $\te$ for these events are in the range 10-60 days, similar 
to those in the MACHO sample.
The optical depth associated with the four events seen in two colors
is $(1.7\pm0.6)\times10^{-8}$ \citep{tis04}, 
not in contradiction with the limit from
the Bright-Sample analysis.  
The 95\% CL limit on the halo fraction is 12\% at $M=0.4M_\odot$.
We prefer to use the limit from the Bright Sample
for the reasons already mentioned:  better understanding of blending and
a superior photometry allowing better rejection of variable stars.
Indeed, inspection of the light curves of EROS2-LMC-8-12 in 
Figures \ref{lm055fig}-\ref{lm085fig} indicates that most are not
especially convincing candidates.  An exception is 
EROS2-LMC-8 but its position in the color-magnitude diagram 
and strongly chromatic magnification indicate that
the event is most likely due to a lens in the Milky Way disk.

\section{Discussion of the results}
\label{discussionsec}

The limits shown in Figure \ref{nevtsfig} rule out machos as the
majority  of Galactic dark matter over the range 
$0.6\times10^{-7}M_\odot<M<15M_\odot$.  
The limits are $f<0.04$ for $10^{-3}M_\odot<M<10^{-1}M_\odot$
and            $f<0.1$ for $10^{-6}M_\odot<M<1M_\odot$.
We note that even stronger limits in the range 
$10^{-7}M_\odot<M<10^{-3}M_\odot$ were found \citep{erosmacho} by
combining the data from the EROS-1 CCD program with
those of the MACHO program.  
These data gave $f<0.07$ for $M\sim3\times10^{-6}M_{\odot}$.
It should also be possible to improve limits in the range $10M_\odot<M<100M_\odot$
by combining the results presented here with the MACHO high mass results
\citep{machobigmass}.
This may narrow the small remaining allowed macho mass range
between the range excluded by microlensing and that excluded
by the abundance of halo wide binary stars \citep{widebinaries}.

Initially,
the EROS and MACHO programs were primarily motivated by the
search for halo brown dwarfs of mass $0.02-0.08M_\odot$.  Such objects
are clearly ruled out as primary components of the Milky Way halo
by Figure \ref{nevtsfig} (implying $f<0.04$) and
the data of the MACHO collaboration \citep{macho57}.

The observation of 17 events by MACHO with $\langle\te\rangle\sim40\,\day$
suggested the existence of machos of $M\sim0.4M_\odot$.
Such objects could be invisible  (e.g. primordial black holes) or
faint (e.g. cool white dwarfs). However, the latter
are not seen at the corresponding level 
in multi-color surveys \citep{gates}
and high
proper-motion surveys \citep[][and references therein]{propermotion,reid}.

At any rate,
the EROS limit (\ref{eroslimit}) is significantly less than the MACHO result
$\tau_{\rm lmc}=1.2^{+0.4}_{-0.3}\times10^{-7}$.
However,
there are considerable differences between the EROS and MACHO  
data sets that may 
help to resolve
the conflict.  Generally speaking, MACHO uses faint stars in dense fields
($1.1\times 10^7$ stars over $13.4\,\deg^2$) while EROS-2 uses bright stars in
sparse fields ($0.7\times 10^7$ stars over $84\,\deg^2$).
Of these bright EROS-2 stars, $0.2\times 10^7$ are in MACHO fields.

The use of dense fields by the MACHO group suggests that the higher MACHO
optical depth may be due, in part, to self-lensing in the inner parts of the
LMC.  
This would contradict  LMC models \citep{jetzerself} 
which suggest that only 1-3 MACHO events should be expected to be 
due to self-lensing.
In fact, MACHO-LMC-14 is known to be
due to self-lensing \citep{lmc14} because  it has a binary source and
the form of the accompanying 
deformation of the lightcurve with respect to the simple
microlensing lightcurve requires that the lens be in the LMC.
A second event, MACHO-LMC-9, is due to a binary lens and the 
self-lensing interpretation
can be avoided only by assuming that
the source is also a binary system 
and that each of the two widely separated components
happened to land on the caustic on the two successive observations
made of the caustic entrance \citep{lmc9}.

If it turns out that the self-lensing rate is higher than
the model estimates 
in the MACHO fields but still negligible in the outer fields of 
the LMC, the disagreement between MACHO and this work is considerably
reduced.  Since only 1/3 of our Bright Sample stars are in MACHO fields,
the EROS-2 95\% CL upper limit on $\tau$ for the MACHO fields is the limit (\ref{eroslimit}) 
multiplied by a factor 3, 
consistent with the \citet{macho57} 
result as modified by \citet{bennetttau}.

A possible 
explanation for the discrepancy that is similar to self-lensing
is the possibility that the events are due to halo lenses but 
the Halo is clumpy and that the MACHO-fields
lie behind a clump of size less than that of the EROS-2 fields.
The effect of a clumpy  halo on the optical depth was discussed
by \citet{clumpyhalo} though they did not discuss directly the
possibility that it could resolve the EROS-MACHO controversy.
At any rate, if this is the cause of the discrepancy, the EROS-2 result
gives the more representative
optical depth because it is based on a larger
solid angle.

The use of faint stars by MACHO may also give an explanation of
the disagreement.  Only two of the 17 MACHO candidates 
(MACHO-LMC-18 and 25)
are sufficiently
bright to be in our Bright-Star Sample.  
Of these two, MACHO-LMC-18 is in the EROS Bright-Sample only because
the EROS starfinder mixed two similar objects that are resolved 
by the MACHO starfinder.

The use of faint stars by MACHO suggests two possible explanations
for the disagreement.
The first would be  contamination
by variable stars that, in our Bright-Star Sample,  
are either not present or identified as such because of superior photometric
precision.
The low photometric precision for the faint LMC stars makes
most of the events less convincing than the events on bright
stars in the Galactic Bulge or the one EROS event
in the SMC.  Indeed, one of the MACHO A events,
MACHO-LMC-23, has been identified as a variable star by a second
variation in the EROS-2 data  shown in Figure \ref{macho23fig} \citep{tis04}.  
As already noted, \citet{bennett23} argued
that MACHO-LMC-23 was, in any case, a weak candidate
and that its variability doesn't call into question the nature
of the other MACHO candidates.  Indeed, some of the MACHO candidates
are very convincing microlensing candidates.
In particular,  MACHO-LMC 1, 5, 9,
14 and 21 are strong candidates based only on MACHO photometry while
\citet{bennett23} argued that MACHO-LMC-4, 13 and 15 are strong
candidates because of high precision followup photometry.  
We might note, however, that of these  candidates,
MACHO-LMC-14 is most likely  due to LMC self-lensing
and MACHO-LMC-5 is due to lensing by a normal red-dwarf
star in the disk of the Milky Way \citep{machomacdo,gouldmacdo}.

The second possible explanation related to the use of faint source stars is 
the large blending effects that must be understood.
\citet{macho57,machoeff} suggest a 30\% systematic error to reflect
this uncertainty.  The experience
with the use of faint stars in the Galactic Bulge suggests that
this uncertainty may be underestimated, though in principle the 
star distribution is better understood in the LMC than in the Bulge.
At any rate, the
extension of the limits presented here, either by EROS or by
OGLE-3 or SuperMACHO,  
will require the use of faint stars and
a good understanding of blending.

\bigskip

\begin{acknowledgements}
Over the last 17 years, the EROS collaboration has profited
from discussions with members of the MACHO collaboration,
especially Charles Alcock (who introduced us to
machos and microlensing), Dave Bennett, Kim Griest and Chris Stubbs.
We thank Annie Robin for help with the Galactic model calculations.
  We are 
  grateful to the technical staff of ESO, La Silla for the support
  given to the EROS-2 project. We thank J-F.~Lecointe and A.~Gomes for the
  assistance with the online computing and
the staff of the CC-IN2P3, especially the team in charge of
the HPSS storage system, for their help with the data management. 
AG was supported by grant AST-0452758 from the NSF and 
JA by the Danish Natural Science Research Council.
Finally, we thank the referee for questions and comments that
led to significant improvements in this paper.

\end{acknowledgements}

\appendix

\section{Candidate selection for the full sample}
\label{fullselectionsec}

\begin{figure*}
\includegraphics[width=.95\textwidth]{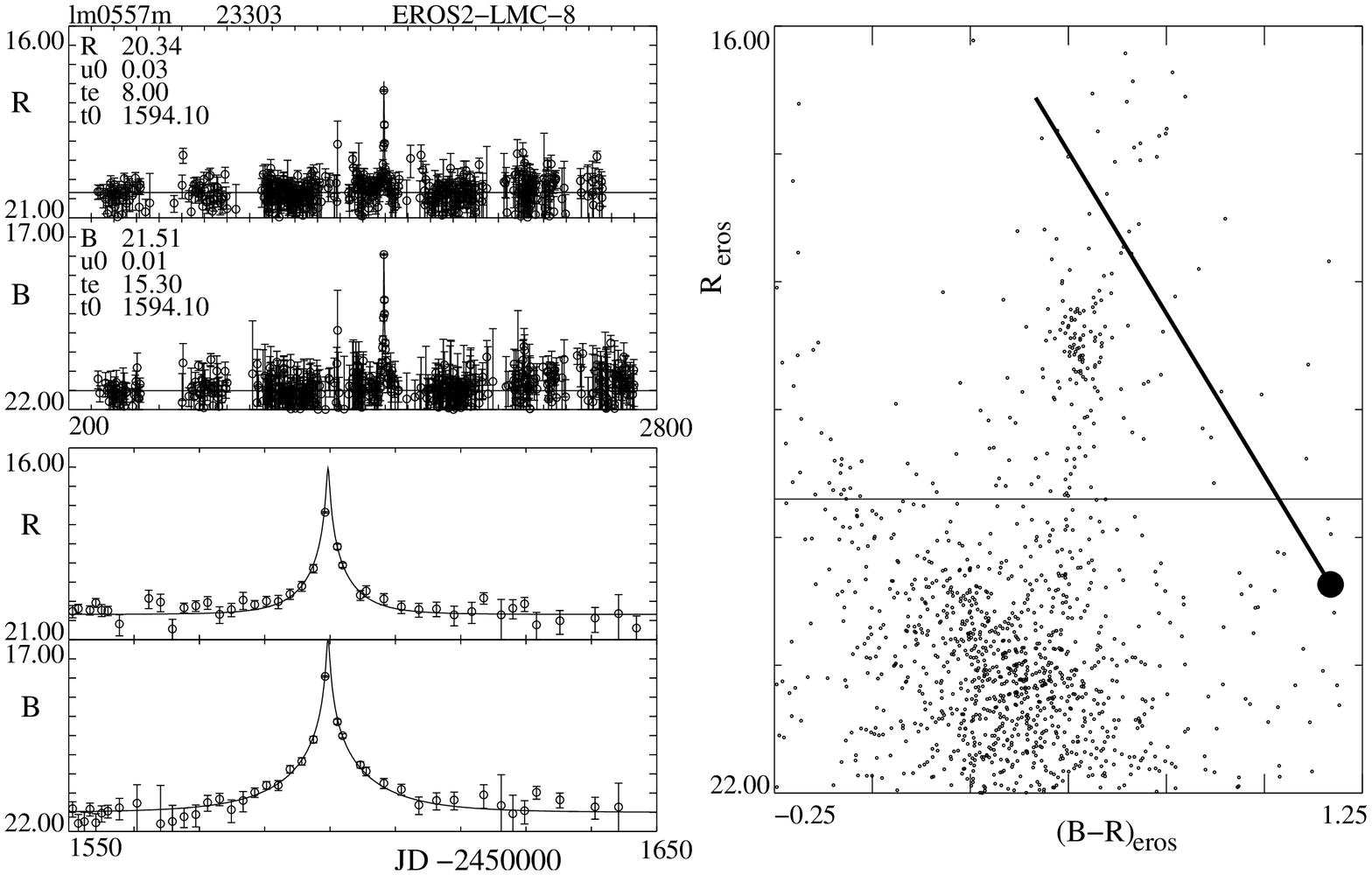}
 \caption{
The light curves of EROS-2 star lm055-7m-23303. 
Also shown is the color-magnitude diagram of the star's CCD-quadrant
and the excursion of the event.}
  \label{lm055fig}
\end{figure*}

\begin{figure*}
\includegraphics[width=.95\textwidth]{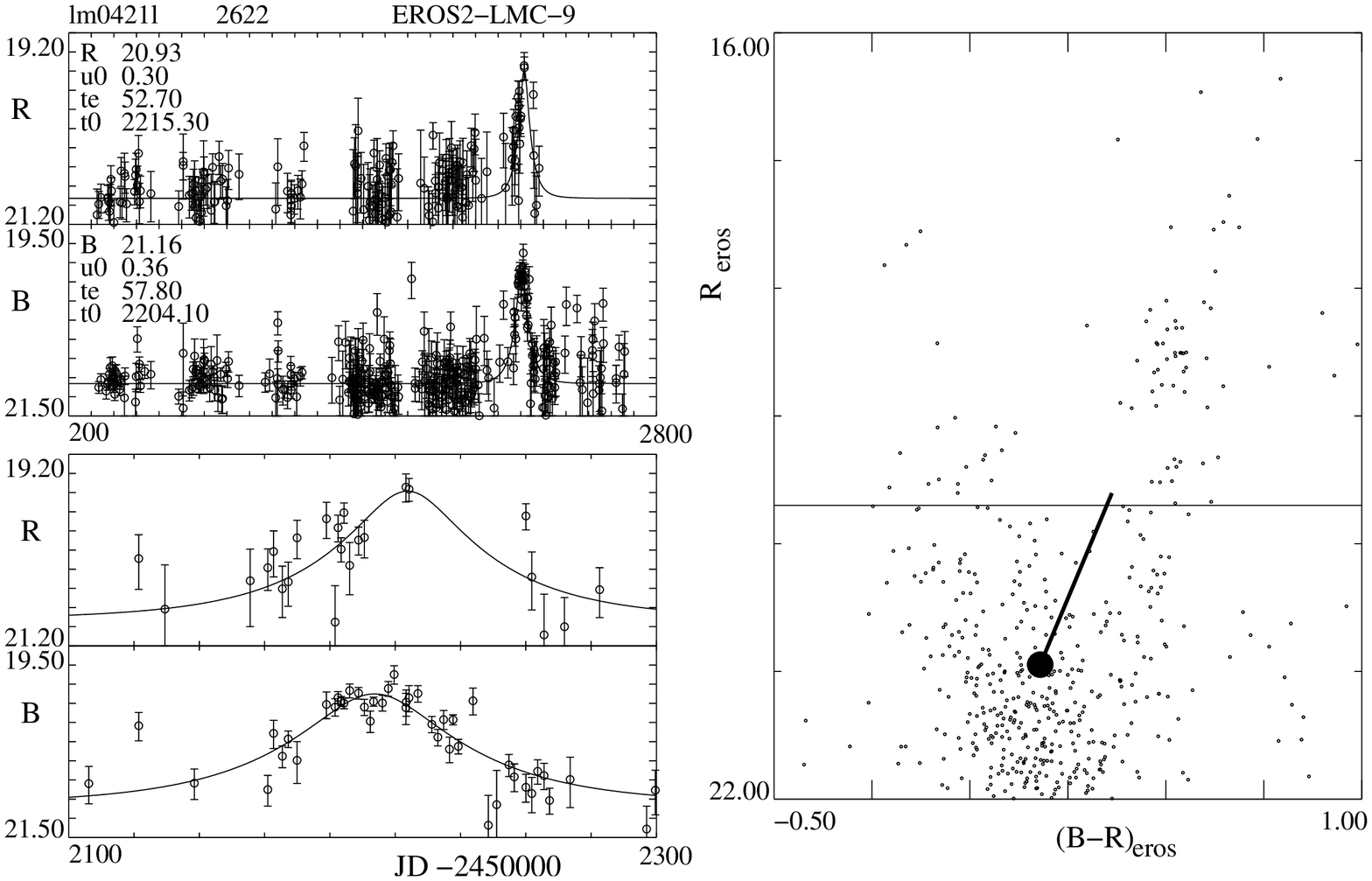}
 \caption{
The light curves of EROS-2 star lm042-1l-2622. 
Also shown is the color-magnitude diagram of the star's CCD-quadrant
and the excursion of the event.}
  \label{lm042fig}
\end{figure*}

\begin{figure*}
\includegraphics[width=.95\textwidth]{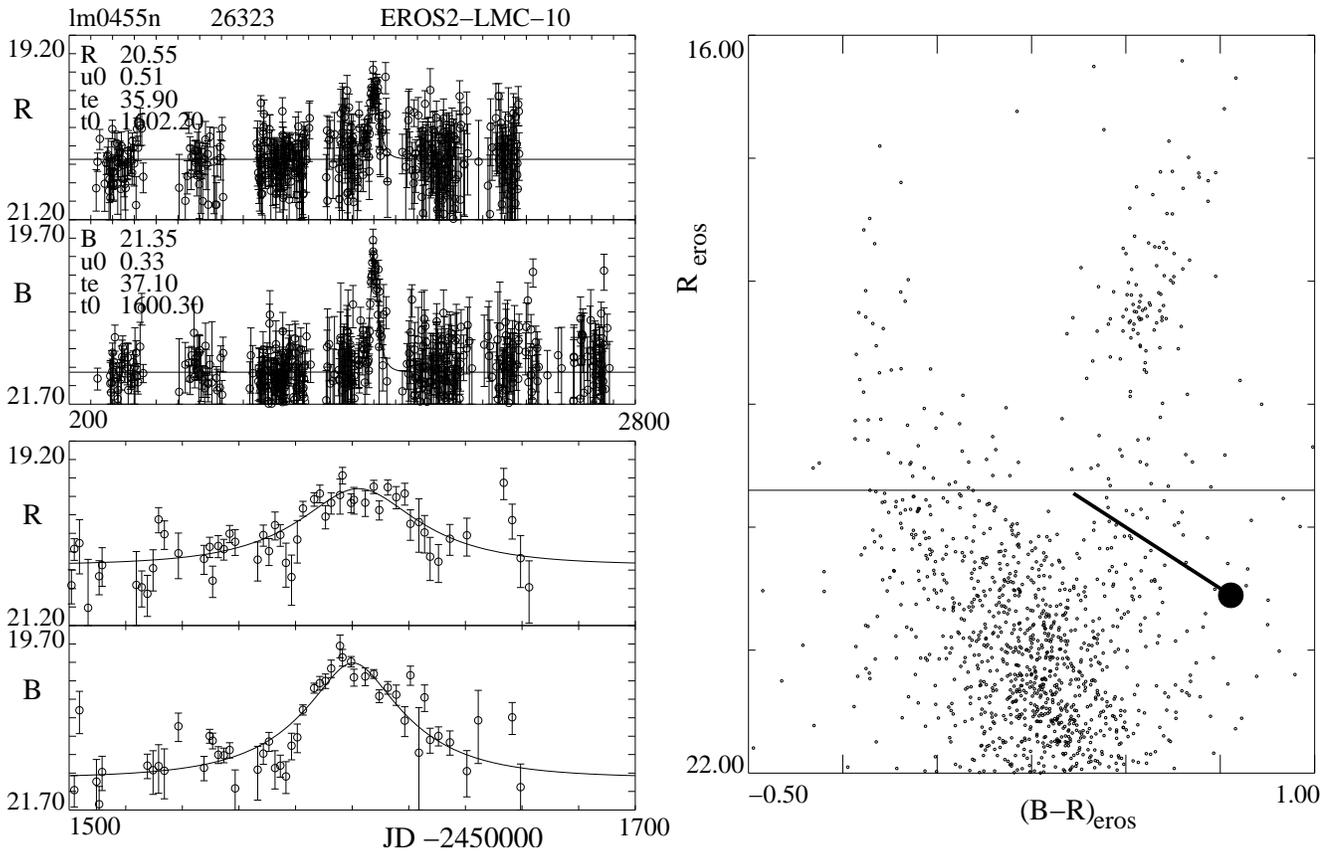}
 \caption{
The light curves of EROS-2 star lm045-5n-26323. 
Also shown is the color-magnitude diagram of the star's CCD-quadrant
and the excursion of the event.
Note the visible asymmetry in the B light curve, possibly indicative of a supernova.}
  \label{lm045fig}
\end{figure*}

\begin{figure*}
\includegraphics[width=.95\textwidth]{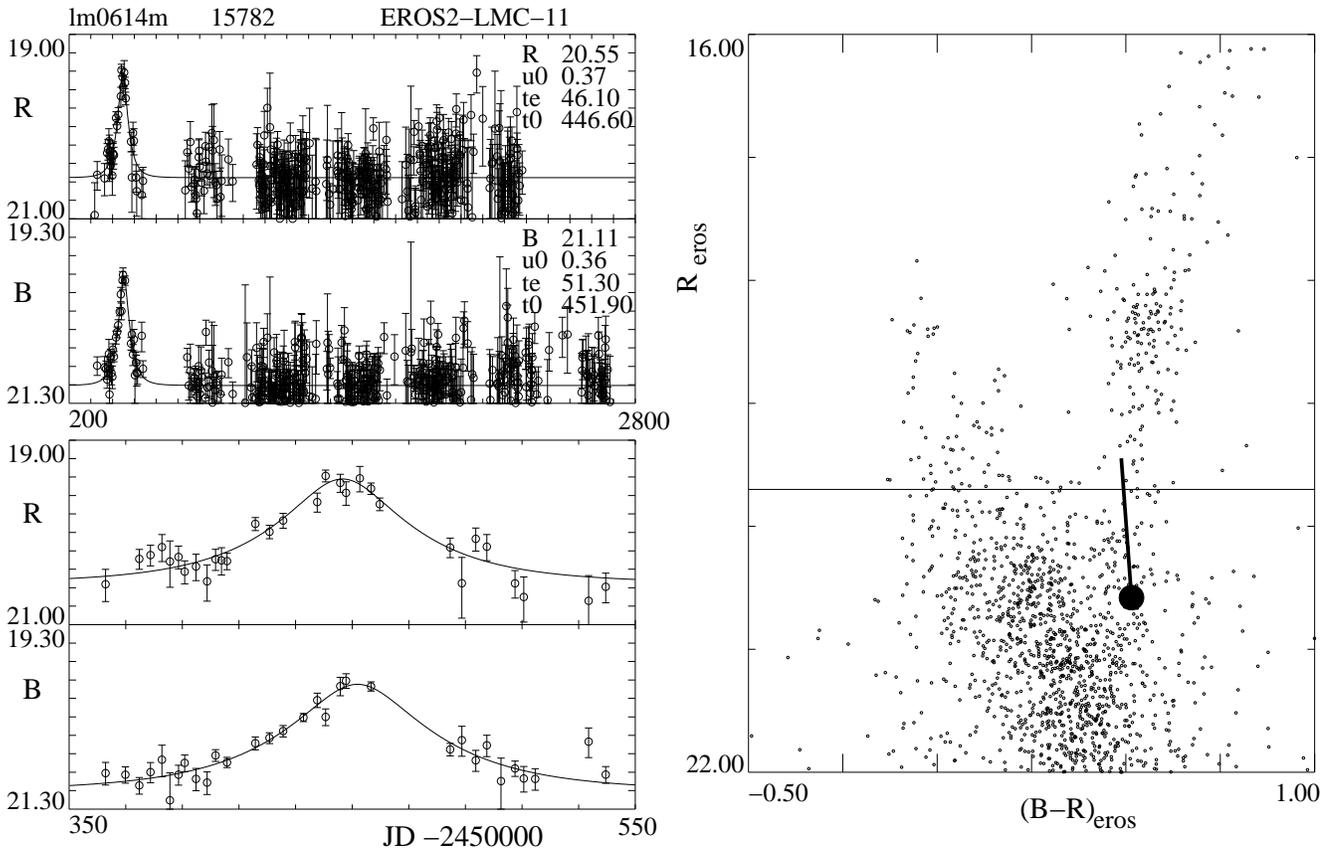}
 \caption{
The light curves of EROS-2 star lm061-4m-15782. 
Also shown is the color-magnitude diagram of the star's CCD-quadrant
and the excursion of the event.}
  \label{lm061fig}
\end{figure*}

\begin{figure*}
\includegraphics[width=.95\textwidth]{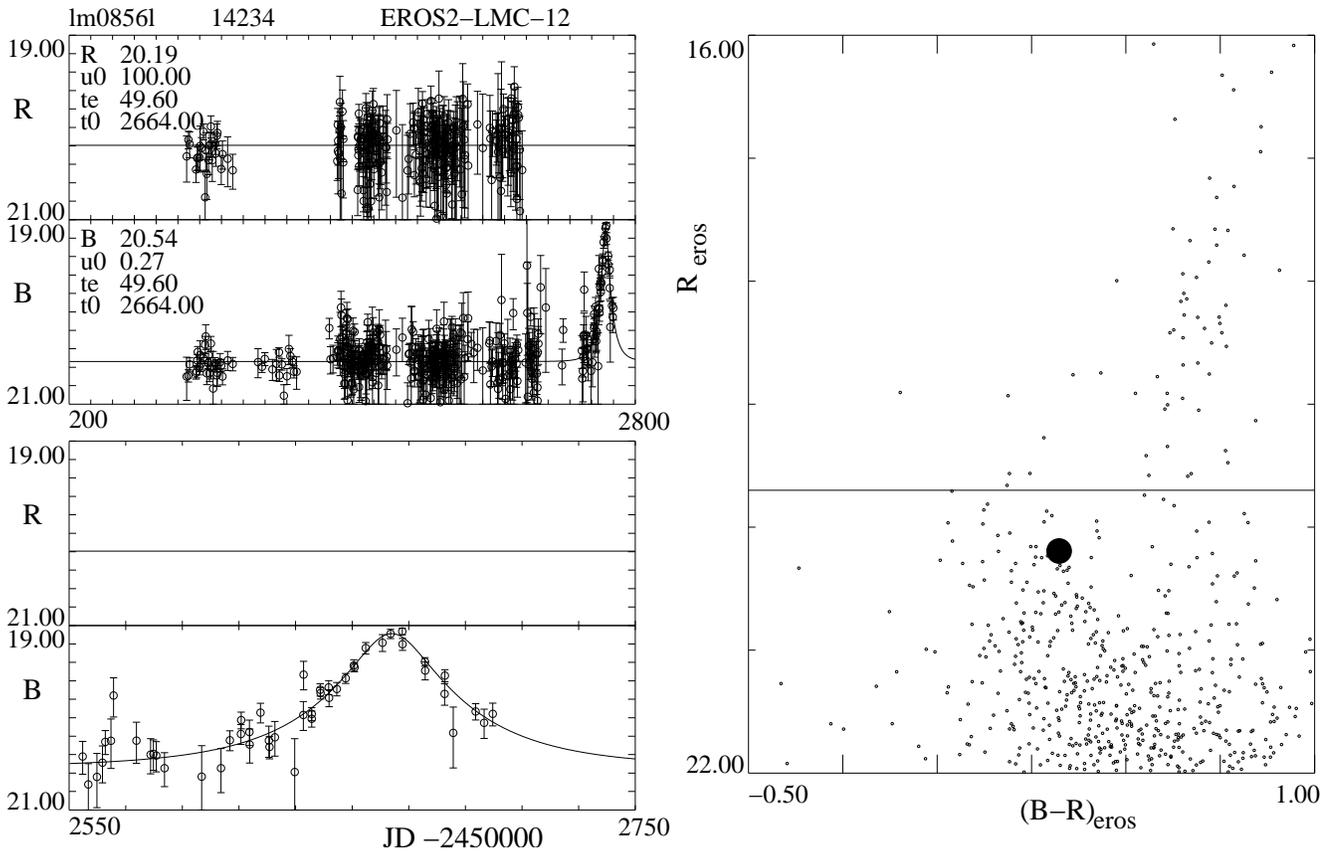}
 \caption{
The light curves of EROS-2 star lm085-6l-14234. 
Also shown is the color-magnitude diagram of the star's CCD-quadrant
and the position of the source's baseline.}
  \label{lm085fig}
\end{figure*}

In this paper we have concentrated on the analysis of the
Bright-Star Sample of stars because this leads to the most reliable
limits of the optical depth. 
A search for microlensing events on the full sample of stars
was also performed \citep{tis04}.  The candidates found in this search
are listed in Table \ref{pateventtable}
and displayed in Figures \ref{lm055fig}-\ref{lm085fig}.  In this section 
we list the selection criteria leading to this set of candidates.

The criteria are very similar to those 
described in Section \ref{analsec} for the Bright-Star Sample.
However, the problems encountered with analyzing low precision
light curves necessarily led to more complicated criteria to
avoid the many spurious events caused by photometric problems.
In all, 17 criteria, $c1-c17$, were applied compared to the 12
criteria, $C1-C12$, applied to the Bright-Star Sample.

The first four criteria are identical to those applied to
the Bright-Star Sample:
\begin{displaymath}
c1\,=\, C1 \hspace*{5mm}
c2\,=\, C2 \hspace*{5mm}
c3\,=\, C3 \hspace*{5mm}
c4\,=\, C4 \;.
\end{displaymath}

We then required that the total amplitude of luminosity variation along the
light curve be greater than 5 times
the point-to-point dispersion, $\sigma_{base}$
in the light curve, recomputed after
excluding the most significant excursions:
\begin{displaymath}
c5: \hspace*{5mm} 
F_{5max}-F_{5min}\,>\,5\sigma_{base} \;.
\end{displaymath} 
Here $F_{5max}$ and $F_{5min}$ are the maximum and minimum fluxes
averaged over any 5 neighboring measurements.

For excursions with a regular variation, the point-to-point dispersion
of the measured fluxes within the excursion,
      $\sigma_{ptp,exc}$,
obtained from the comparison of each measured flux with the linear
interpolation of its two neighbors (in time), is normally smaller than the
global dispersion,
   $\sigma_{exc}$,
i.e. the width of the distribution of all fluxes within the excursion.
To exclude irregular variations, we require that their ratio be sufficiently
small;
\begin{displaymath}
c6: \hspace*{5mm} 
\frac{\sigma_{ptp,exc}}{\sigma_{exc}} \,<\, 0.90 \;.
\end{displaymath}

Cuts $c4-c6$ are applied independently in the two passbands. Light curves are
retained if they are selected in one passband at least. This reduces the
star sample to slightly less than 0.1\% (28,500 objects) of the full sample.

The next two criteria are similar to $C5$ and $C9$ applied to the
Bright-Star Sample
\begin{displaymath}
c7:\hspace*{5mm}  \left[\frac{ \chi^2 - N_{dof} }
{  \sqrt{2 N_{dof} }}\right]_{base}\;
<\; 15 \;,
\end{displaymath}
\begin{displaymath}
c8:\hspace*{5mm}
 \left[\frac{ \chi^2 - N_{dof} }{  \sqrt{2 N_{dof} }}\right]_{peak}\;
<\; 10 \;.
\end{displaymath}

The next cut, $c9$, deals with a background connected to the realignment
of the telescope optics in May 1998, which slightly changed the PSF.
Faint stars near a diffraction feature in the PSF of bright stars were
affected; this was seen as light curves with two plateaus, one 
(higher) before and one (lower) after May 1998. 
A rather complicated algorithm was developed to identify
such light curves. It relies
mostly on the relative flux values of the two stars, their distance and
the relative height of the two plateaus in the light curve. 
The details can be found in \citep{tis04} but we note that it
affects our efficiency only for $\te>150\,\day$.
\begin{displaymath}
c9: \hspace*{5mm}{\rm no\;influence\;of\;telescope\;realignment}
\end{displaymath}

At this point we accepted only light curves that passed criteria
$c1-c9$ in both colors:
\begin{displaymath}
c10:\hspace*{5mm}
c1-c9\; {\rm red\;and\;blue}
\end{displaymath}

Criterion c11 is similar to C7:
\begin{displaymath}
c11:\hspace*{5mm}
N_{peak}>4 \;;{\rm \;\geq2\;points\;in\;rise\;;\;\geq2\;points\;in\;fall\;.}
\end{displaymath}

The remaining cuts follow closely the cuts for the Bright-Star Sample
\begin{displaymath}
c12\;=\;C6
\end{displaymath}
\begin{displaymath}
c13:\hspace*{5mm}
u_{0r}<1 \;\;{\rm or} \;\;u_{0b}<1
\end{displaymath}
\begin{displaymath}
c14:\hspace*{5mm}
\frac{\chi^2_{ct} - \chi^2_{ml}}{\chi^2_{ml} / N_{dof}}
\frac{1}{\sqrt{2 N_{dof,peak}}} \,>50 \;\;{\rm in} \;R\;{\rm or}\;B \;.
\end{displaymath}
Criterion $c15$ eliminates light echos from SN1987a by requiring
\begin{displaymath}
c15:\hspace*{5mm}
{\rm Distance\;event-SN1987a}>30\,{\rm arcmin} \;.
\end{displaymath}
Finally, $c16$ and $c17$ are the blue bumper and supernovae  criteria
$C10$ and $C11$ supplemented by a visual rejection of events
superimposed on background galaxies.

The first six events in Table \ref{pateventtable} passed the
cuts $c1$-$c17$.  A monochromatic analysis was performed where
$c10$ was removed and candidates were only required to pass the
other criteria in either of the two colors.  About 500 events
were scanned by eye.  Only one event, EROS2-LMC-12,
was found, the others being due to photometric problems or long period 
variable stars.

\bibliographystyle{aa}

\begin{thebibliography}{}

\bibitem[Afonso et al.(1998)]{erossmcdouble}  Afonso, C., Albert, J. N., Alard, C. et al. (EROS-2 coll.), 1998, \aap, 337, L17

\bibitem[Afonso et al.(1999)]{smc2}  Afonso, C., Alard, C., Albert, J.N. et al. (EROS-2 coll.), 1999, \aap, 344, L63

\bibitem[Afonso et al.(2000)]{smcdouble} Afonso C., Alard, C., Albert, J.N. et al. (EROS-2 , MACHO -GMAN , MPS , OGLE and PLANET colls.),  2000, \apj, 532, 340

\bibitem[Afonso et al.(2003{\natexlab{a}})]{EROSMC} Afonso, C., Albert, J. N., Andersen, J., et al. (EROS-2 coll.) 2003a, \aap, 400, 951 

\bibitem[Afonso et al.(2003{\natexlab{b}})]{Afonso} Afonso, Albert, J. N., Alard, C., et al. (EROS-2 coll.) 2003b, \aap, 404, 145




\bibitem[Alcock et al.(1997a)]{alc97a}   Alcock, C. Allsman, R.A., Alves, D. et al. (MACHO coll.), 1997a, \apj, 486, 697
\bibitem[Alcock et al.(1997b)]{smc1macho} Alcock, C. Allsman, R.A., Alves, D. et al. (MACHO coll.), 1997b, \apj, 491, L11

\bibitem[Alcock et al.(1998)]{erosmacho} Alcock, C. Allsman, R.A., Alves, D. et al. (MACHO and EROS-1 colls.), 1998, \apj,  499, L9

\bibitem[Alcock et al.(2000a)]{lmc9}  Alcock, C., Allsman R.~A., Alves D.~R., et al. (MACHO coll.) 2000a, \apj, 541, 270


\bibitem[Alcock et al.(2000b)]{macho57}Alcock, C., Allsman, R. A., Alves, D. R., et al. (MACHO coll.) 2000b, \apj, 542, 281

\bibitem[Alcock et al.(2001a)]{machoeff}Alcock, C., Allsman, R. A., Alves, D. R., et al. (MACHO coll.) 2001a, \apjs, 136, 439
\bibitem[Alcock et al.(2001b)]{machobigmass}Alcock, C., Allsman, R. A., Alves, D. R., et al. (MACHO coll.) 2001b, \apj, 550, L169
\bibitem[Alcock et al.(2001c)]{lmc14}Alcock, C., Allsman, R. A., Alves, D. R., et al. (MACHO coll.) 2001c, \apj, 552, 259
\bibitem[Alcock et al.(2001d)]{machomacdo}Alcock, C., Allsman, R. A., Alves, D. R., et al. (MACHO coll.) 2001d, \nat, 414, 617

\bibitem[Ansari et al.(1996a)]{erosplates}Ansari, R., Cavalier, F., Moniez, M., et al. (EROS-1 coll.) 1996a, \aap, 314, 94
\bibitem[Ansari et al.(1996b)]{ANS96}Ansari, R. et al. (EROS-2 coll.) 1996b, Vistas in Astronomy, 40, 519

\bibitem[Assef et al.(2006)]{parsmc}Assef, R.J., Gould, A., Afonso, C., et al 2006, \apj,  649, 954

\bibitem[Aubourg et al.(1993)]{aub93} Aubourg, {\'E}. et al., (EROS-1 coll.), 1993, \nat, 365, 623

\bibitem[Bauer et al.(1997)]{bau97}Bauer, F. et al. (EROS-2 coll.) 1997,
Proceeding of the ``Optical Detectors for Astronomy'' workshop, ESO

\bibitem[Beaulieu et al.(1995)]{beau95} Beaulieu, J.-Ph., Ferlet, R., Grison, Ph. et al., (EROS-1 coll.), 1995, \aap, 299, 168

\bibitem[Bennett et al.(2005)]{bennett23}Bennett, D. P., Becker, A. C., \& Tomaney, A. 2005, \apj, 631, 301

\bibitem[Bennett(2005)]{bennetttau}Bennett, D. P. 2005, \apj, 633, 906



\bibitem[Bissantz \& Gerhard(2002)]{Bissantz} Bissantz, N. and Gerhard, O., 2002,       \mnras, 330, 591

\bibitem[Bond et al.(2002)]{macho99lmc2}Bond, I.A., Rattenbury, N.J., Skuljan, J., et al. 2002, \mnras, 333, 71-83.

\bibitem[Calchi Novati et al.(2005)]{PH2005} Calchi Novati, S., Paulin-Henriksson, S., An, J.,  et al. (AGAPE coll.) 2005,            \aap, 443, 911 

\bibitem[de Jong  et al.(2006)]{mega2005} de Jong, J. T. A., Widrow, L.M., Cseresnjes, P.,  et al. (MEGA coll.) 2006, \aap, 446, 855

    
\bibitem[Evans \& Belokurov(2002)]{EVA02}Evans, N.W. \&  Belokurov, V.  2002,                \apj, 567,  L119

\bibitem[Fellhauer et al.(2006)]{fellhauer}Fellhauer, M., Belokurov, V., Evans, N.W., et al.  2006, \apj, 651, 167

\bibitem[Gates et al.(2004)]{gates} Gates, E., Gyuk, G., Harris, H. C., et al. (SDSS coll.) 2004, \apj, 612, 132


\bibitem[Goldman et al.(2002)]{propermotion}Goldman, B., Afonso, C., Alard, C., et al. (EROS-2 coll.)  2002,            \aap, 389, 69

\bibitem[Gould(2004)]{gouldmacdo}Gould, A. 2004,            \apj, 606, 319

\bibitem[Graff \& Gardiner(1999)]{graffsmc}Graff, D. \& Gardiner, L.T. 1999,            \mnras, 307, 577

\bibitem[Griest(1991)]{GRI91}Griest, K.  1991,              \apj, 366, 412 


\bibitem[Hamadache et al.(2006)]{hamaa}Hamadache, C., Le Guillou, L., Tisserand, P.,  et al  (EROS-2 coll.) 2006, \aap,  454, 185


\bibitem[Han \& Gould(2003)]{HanGould03} Han, C. and Gould, A. 2003,           \apj, 592, 172

\bibitem[Holopainen et al.(2006)]{clumpyhalo}Holopainen, J., Flynn, C., Knebe, A., et al., 2006, \mnras, 368, 1209


\bibitem[Joshi et al.(2005)]{nainital}Joshi, Y.C., Pandey, A.K., Narasimha, D. \& Sagar, R. (Nainital coll.) 2005, \aap, 433, 787


\bibitem[Lasserre(2000)]{lasthesis}Lasserre, T. 2000,  Thesis Universit\'e de Paris VI, {\it http://tel.ccsd.cnrs.fr/} 

\bibitem[Lasserre et al.(2000)]{las00} Lasserre, T., Afonso, C., Albert, J.N.et al., (EROS-2 coll.), 2000, \aap, 355, L39

\bibitem[Le Guillou(2003)]{triton} Le Guillou, L. 2003, Thesis Universit\'e de Paris VI, {\it http://tel.ccsd.cnrs.fr/}

\bibitem[Mancini et al.(2004)]{jetzerself}Mancini, L., Calchi Novati, S., Jetzer, Ph., \& Scarpetta, G. 2004, \aap, 427, 61

\bibitem[Milsztajn et al.(2001)]{mil00}  Milsztajn, A. and Lasserre, T. (for the EROS-2 coll.), 2001, Nucl. Phys. B  (Proc. Suppl) 91, 413


\bibitem[Paczy\'nski(1986)]{Pac1986}Paczy\'nski, B. 1986, \apj, 304, 1


\bibitem[Paczy{\'n}ski(1996)]{Pac96}  Paczy{\'n}ski, B., 1996, ARA\&A, 34, 419

\bibitem[Palanque-Delabrouille et al.(1998)]{smc1}   Palanque-Delabrouille, N., Afonso, C., Albert, J. N., et al. (EROS-2 coll.), 1998, \aap, 332, 1

\bibitem[Petrou(1981)]{Pet81}   Petrou, M., 1981. Ph.D. thesis, University of Cambridge

 
\bibitem[Popowski et al.(2005)]{machocgpop}Popowski, P., Griest, K., Thomas, C. L., et al. (MACHO coll.) 2005, \apj, 631, 879

\bibitem[Reid(2005)]{reid}Reid, I.N. 2005, ARA\&A, 43, 247

\bibitem[Renault et al.(1997)]{renault97}Renault, C., Afonso, C., Aubourg, E., et al. (EROS-1 coll.) 1997, \aap, 324, L69
\bibitem[Renault et al.(1998)]{renault98}Renault, C., Aubourg, E., Bareyre, P., et al. (EROS-1 coll.)  1998, \aap, 329, 522


\bibitem[Riffeser et al.(2003)]{Riff03}Riffeser, A., Fliri, J., Bender, R., et al. (WeCapp coll.) 2003,     \apjl, 599, L17

\bibitem[Robin et al.(2003)]{besancon}Robin, A., Reyl\'e, C., Derri\`ere, S., \& Picaud, S.  2003,     \aap, 409, 523; 
and  2004, \aap, 416, 157

\bibitem[Sackett and Gould(1993)]{sakgould}Sackett, P.D. and Gould, A., 1993, \apj, 419, 648


\bibitem[Sumi et al.(2006)]{sumiogle}Sumi, T., Wozniak, P. R.,  Udalski, A., et al. (OGLE coll.) 2006, \apj, 636, 240.

\bibitem[Tisserand(2004)]{tis04}Tisserand, P. 2004,  Thesis Universit\'e de Nice, {\it http://tel.ccsd.cnrs.fr/} 


\bibitem[Udalski et al,(1997)]{uda97}   Udalski, A., Szyma\'nsi, M., Kumiak, M., et al. (OGLE coll.), 1997, Acta Astron., 47, 431

\bibitem[Udalski et al.(2000a)]{UDA00}Udalski A., Zebrun K., Szymanski M., et al. (OGLE coll.) 2000a, Acta Astron., 50, 1

\bibitem[Udalski et al.(2000b)]{ogleIIphotometry}Udalski, A., Szymanski, M., Kubiak, M., et al. (OGLE coll.) 2000b, Acta Astron., 50, 307


\bibitem[Uglesich et al.(2004)]{Ugle04}Uglesich, R.R., Crotts, A. P. S., Baltz, E.A., et al. (VATT coll.) 2004, \apj, 612, 877

\bibitem[Wood and Mao(2005)]{woodmao}Wood, A. and Mao, S. 2005, \mnras,  362, 945


\bibitem[Yoo, Chanam\'e and Gould(2004)]{widebinaries}Yoo, J., Chanam\'e J., and Gould, A.  2004,  \apj, 601, 311

\bibitem[Zaritsky et al.(2004)]{zaritsky}Zaritsky Z., Harris J., Thompson I., and Grebel, E.  2004,  \aj, 128, 1606

\end{thebibliography}

\end{document}